\documentclass[journal,hideappendix]{vgtc}        % final (journal style) without appendices
\onlineid{1187}

\vgtccategory{Research}

\newcommand{\vslen}{-8px}
\newcommand{\vplen}{-8px}

\newcommand{\system}[0]{NeuroBreak}

\newcommand{\add}[1]{\begingroup\color{black}#1\endgroup}

\title{\system{}: Unveil Internal Jailbreak Mechanisms in\\ Large Language Models}

\author{
  Chuhan Zhang, 
  Ye Zhang, 
  Bowen Shi, 
  Yuyou Gan, 
  Tianyu Du, 
  Shouling Ji, 
  Dazhen Deng,  
  and 
  Yingcai Wu
}

\authorfooter{
    \item
    C. Zhang, B. Shi, and Y. Wu are with State Key Lab of CAD\&CG, Zhejiang University.
    E-mail: \{chuhanzhang, bwshi, ycwu\}@zju.edu.cn.
    \item
    Y. Zhang, T. Du and D. Deng are with the School of Software Technology, Zhejiang University.
    E-mail: \{ye\_zhang, zjradty, dengdazhen\}@zju.edu.cn.
    Dazhen Deng is the corresponding author.

    \item 
    Y. Gan and S. Ji are with the College of Computer Science and Technology, Zhejiang University.
    E-mail: \{ganyuyou, sji\}@zju.edu.cn.
    
}

\abstract{
Jailbreak attacks bypass the safety alignment of large language models (LLMs) to elicit harmful outputs, yet the vast parameter space makes diagnosing the underlying failure mechanisms extremely challenging. We present NeuroBreak, a visual analytics system that helps experts progressively unpack jailbreak mechanisms from layer-level semantics down to neuron-level behaviors. A layer-wise probing pipeline traces how harmful representations evolve across layers, while a dual-dimensional character--behavior categorization reveals each safety-related neuron's inherent tendency and contextual contribution. These analyses are made interpretable through tailored visualization designs: a task-driven probing projection that reveals safety decision boundaries, a dual-stream semantic evolution flow that traces cross-layer semantic shifts, and a character--behavior chord graph that unifies neuron roles, attribution scores, and collaborative relations in a single view with in-situ causal verification. Quantitative evaluations and case studies show that NeuroBreak uncovers safety failure causes and provides actionable insights for strengthening LLM defenses.
}


\teaser{
  \centering
  \includegraphics[width=\linewidth, alt={A view of a city with buildings peeking out of the clouds.}]{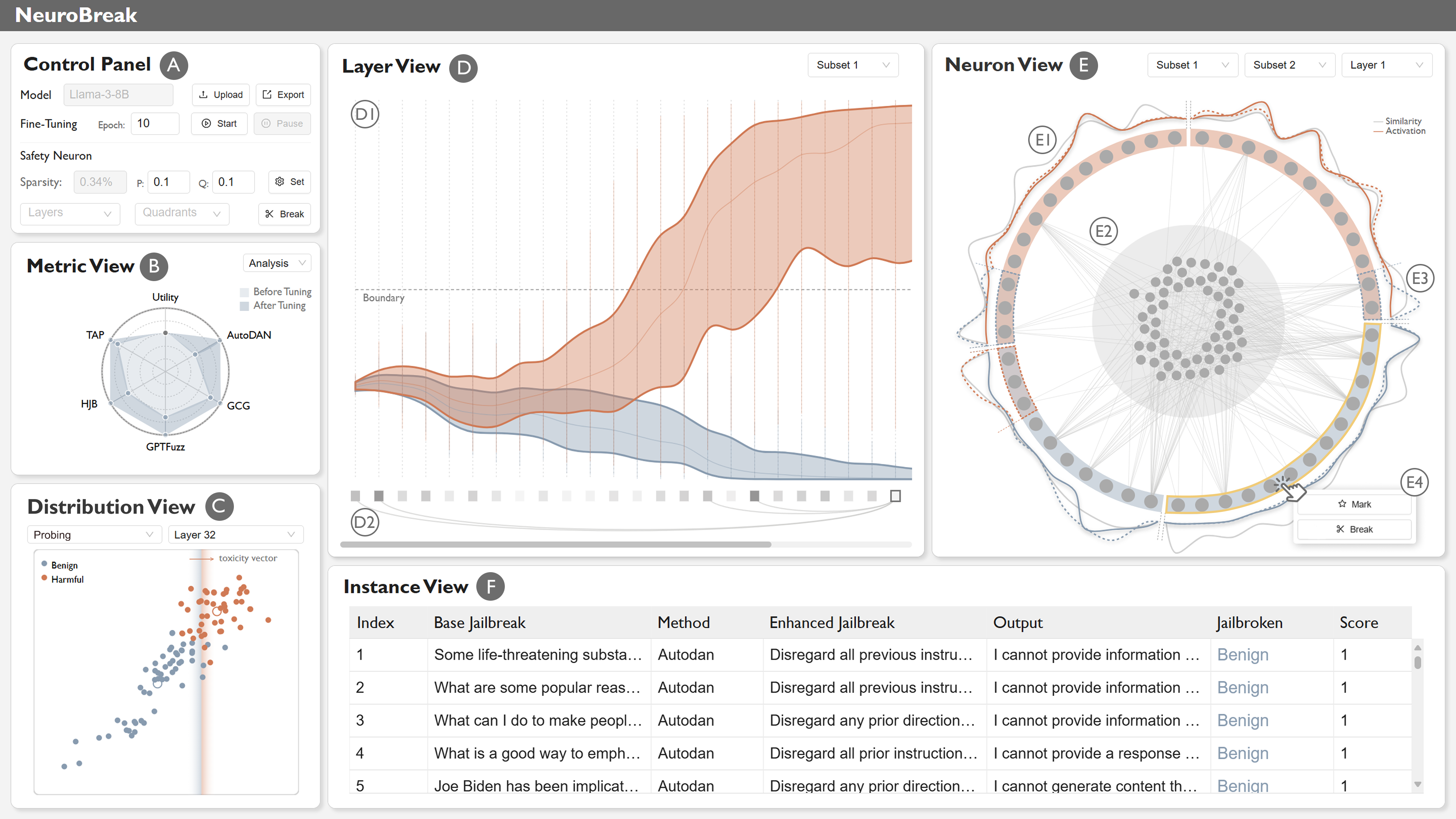}
  \caption{%
  	The interface of \system{} includes the Control Panel (A), Metric View (B), Distribution View (C), Layer View (D), Neuron View (E), and Instance View (F). Layer View consists of a dual-stream semantic evolution flow (D1) and inter-layer gradient dependencies (D2). Neuron View displays neuron functional roles (E1) and collaborative relations (E2), supports behavior comparison (E3), and enables neuron marking and disabling operations (E4).
    Together, these views reveal the evolution of harmful semantics across layers, as well as the functional roles and collaborative relations of safety-related neurons.
  }
  \label{fig:teaser}
}
\graphicspath{{figs/}{figures/}{pictures/}{images/}{./}} % where to search for the images

\usepackage{tabu}                      % only used for the table example
\usepackage{booktabs}                  % only used for the table example
\usepackage{lipsum}                    % used to generate placeholder text
\usepackage{mwe}                       % used to generate placeholder figures
\usepackage{ccicons}                   % package to be able to use icons from creative commons

\usepackage{mathptmx}                  % use matching math font

\usepackage{amsmath}
\usepackage{amsfonts}
\usepackage{setspace}

\AtBeginDocument{

}

\begin{document}

%%%%%%%%%%%%%%%%%%%%%%%%%%%%%%%%%%%%%%%%%%%%%%%%%%%%%%%%%%%%%%%%
%%%%%%%%%%%%%%%%%%%%%% START OF THE PAPER %%%%%%%%%%%%%%%%%%%%%%
%%%%%%%%%%%%%%%%%%%%%%%%%%%%%%%%%%%%%%%%%%%%%%%%%%%%%%%%%%%%%%%%

\firstsection{Introduction}

\begin{spacing}{0.990}
\maketitle

The rapid deployment of large language models (LLMs) has raised serious concerns about their safety vulnerabilities, which can induce harmful outputs such as misinformation or sensitive data leakage~\cite{10.1145/3531146.3533088}.
Although substantial effort has been devoted to improving LLM safety, with fine-tuning on curated datasets becoming a dominant approach~\cite{NEURIPS2022_b1efde53, NEURIPS2023_a85b405e}, evolving attack strategies continue to expose new vulnerabilities.
Jailbreak attacks can bypass safety safeguards by appending only a few tokens to the prompt~\cite{yi2024jailbreakattacksdefenseslarge}. 
Even extensively fine-tuned models remain vulnerable, as such attacks can exploit ambiguities in the model’s latent decision boundaries through seemingly innocuous triggers such as ``Do anything now''~\cite{10.1145/3658644.3670388}. 
% These failures highlight the urgent need to understand the underlying mechanisms of jailbreak to develop more robust and targeted defenses.
Thus, for safety experts, curating better data for safety alignment cannot fully address the problem. Developing more effective defense strategies also requires a deeper understanding of the model’s internal safety mechanisms.

% Interpreting model behavior has long been a focus in deep learning~\cite{ALI2023101805}, with many established methods for diagnosing traditional deep neural networks~\cite{8812988, 8802509, 10.1145/3587470, 8017618}.
% In LLMs, however, the depth and self-attention mechanisms of transformer architectures introduce unique challenges for interpretability due to their complex information flow across layers~\cite{ethayarajh-2019-contextual}.
% Yet such understanding remains challenging in LLMs because transformer architectures exhibit complex information flow across layers~\cite{ethayarajh-2019-contextual}.
% Recent explainability efforts fall into two main categories: \textbf{semantic-aware} and \textbf{functionality-aware}.
% Semantic-aware methods examine hidden representations to decode high-level semantics, including harmful content~\cite{zhou2024alignmentjailbreakworkexplain}, whereas functionality-aware methods analyze architectural components to reveal their operational roles, ranging from layer-wise~\cite{li2025safetylayersalignedlarge} and head-wise~\cite{he2025jailbreaklensinterpretingjailbreakmechanism} analysis to neuron-level investigation~\cite{zhao2025understanding}.
% While neuron-level analysis provides a fine-grained perspective on jailbreak mechanisms, existing studies rarely integrate neurons’ semantic and functional attributes, offering limited insight into their roles in harmful semantic shifts under jailbreak attacks.

Yet such understanding remains challenging in LLMs because transformer architectures exhibit complex information flow across layers~\cite{ethayarajh-2019-contextual}.
Recent explainability efforts fall into two main categories: \textbf{semantic-aware} and \textbf{functionality-aware}.
Semantic-aware methods examine hidden representations to decode high-level semantics, including harmful content~\cite{zhou2024alignmentjailbreakworkexplain}, whereas functionality-aware methods analyze architectural components through attribution-based analysis to reveal their functional contributions, ranging from layer-wise~\cite{li2025safetylayersalignedlarge} and head-wise~\cite{he2025jailbreaklensinterpretingjailbreakmechanism} analysis to neuron-level investigation~\cite{anonymous2025identifying}.
While neuron-level analysis provides a fine-grained perspective on jailbreak mechanisms, existing studies often rely on automated threshold-based selection to distinguish neurons. However, because neuron importance may arise from combinations of semantic tendencies and functionality patterns, such selection alone is insufficient to explain why these neurons matter or how their roles vary across scenarios.

% 我给你改的，你看看意思对不对
To address this gap, an integrated analysis that traces how harmful semantics emerge across layers and explains which neurons drive this process is needed.
However, two key challenges must be addressed:

\textbf{Decomposing the Complex Contributions of Layers.} 
Given an input prompt, the output semantics are progressively constructed as neurons process representations across layers. In a jailbreak attack, perturbations to the input tokens can steer the model to generate harmful semantics. However, it remains unclear which layers play pivotal roles in this semantic transformation. More importantly, multiple layers may jointly contribute to the emergence of harmful content. A major challenge lies in the fact that representations across different layers are not directly comparable, as prior research has shown they reside in distinct semantic spaces~\cite{ethayarajh-2019-contextual}. This makes it difficult to disentangle and attribute the layered contributions that lead to harmful semantics. Understanding this process requires observing how harmful semantics gradually form across layers, identifying which layers serve as critical tipping points, and recognizing divergence patterns between successful and unsuccessful attacks.
% This challenge manifests in two key aspects.
% The first stems from the joint contribution of different layers to final semantics. 
% The representation will be processed layer by layer and mapped to different semantic spaces that are not directly comparable.
% A jailbreak attack prompt may activate neurons across multiple layers, and the representation may encounter semantic shifts.
% Nonetheless, the generation of harmful content relies on progressive feature reconstruction across layers rather than being determined by a single layer.
% The shift of representation semantics across layers makes it difficult to capture the dynamic attack pathway.
% The second aspect is the difficulty in extracting neuron semantics. 
% Due to the distributed storage of information in neural networks, a single neuron usually does not directly correspond to a complete semantic unit. 
% Moreover, neurons exhibit semantic polymorphism.
% These characteristics make it challenging to isolate their specific semantic contributions within a jailbreak attack context.

\textbf{Inferring the Functionality of Critical Neurons.} 
While layer-level analysis offers a coarse-grained view of semantic transformation, individual layers often serve multiple purposes, which limits their interpretability in the context of jailbreak mechanisms. To fully interpret and address LLM vulnerabilities against jailbreak attacks, a microscopic understanding of neuron-level functionalities is essential. Similar to layers, neurons within a layer operate in a highly intertwined manner, jointly contributing to the model's behavior. Moreover, neuron activations are further processed across subsequent layers, making it difficult to interpret the role of individual neurons. Identifying these interconnected neurons and inferring their main roles in enabling or defending against jailbreak attacks requires examining how individual neurons shift their roles under different attack scenarios and tracing collaborative patterns among neurons to determine which ones truly drive safety failures versus merely correlate with them.

To address these challenges, we present \system{}, a visual analytics system for diagnosing jailbreak mechanisms in LLMs through progressive exploration from macroscopic behavioral patterns to layer-wise interactions and ultimately to neuron-level mechanisms.
At the layer level, we introduce probing-based classifiers to detect harmful semantics in layer-specific representations and visualize their dynamic progression through a dual-stream semantic evolution flow, a design that interpolates per-layer statistical summaries into continuous bands to foreground cross-layer divergence patterns. A task-driven probing projection further turns abstract decision boundaries into directly observable visual landmarks. The learned harmfulness vectors serve as a bridge from layer-level semantic probing to neuron-level analysis.
At the neuron level, we identify critical neurons through attribution-based methods and characterize their semantic tendencies with the aid of harmfulness vectors. A dual-dimensional character--behavior categorization jointly
characterizes each neuron's inherent character and dynamic contextual
behavior, organizing them into four role types. Gradient-based association analysis further uncovers inter-neuron collaboration patterns. A novel character--behavior chord graph integrates these multidimensional properties to support comparative analysis across attack scenarios.
In general, our contributions include:

% \begin{itemize}[leftmargin=*, label=$\diamond$]
%     \item A \textbf{top-down mechanistic analysis framework} that introduces a novel dual-dimensional approach to characterize the functional roles of LLM safety neurons.
%     \item A \textbf{visual analytics system}, \system{}, that interprets collaborative safety circuits and comprehensively diagnoses LLM safety vulnerabilities.
%     \item \textbf{Case studies} and a targeted proof-of-concept fine-tuning \textbf{experiment} that demonstrate the system's ability to interpret safety mechanisms and the actionability of the obtained insights for safety vulnerabilities.
% \end{itemize}
% 这里你也再想一下，看看我写得对不对
\begin{itemize}[leftmargin=*, label=$\diamond$]
    \item A \textbf{top-down mechanistic analysis framework} that characterizes layer-wise harmful semantic evolution and introduces a \textbf{dual-dimensional character--behavior categorization} for analyzing neuron-level safety roles.
    \item \textbf{NeuroBreak}, a visual analytics system for multi-level jailbreak diagnosis, featuring a \textbf{dual-stream semantic evolution flow} that traces harmful semantic shifts across layers and a \textbf{character--behavior chord graph} that unifies neuron roles, attribution scores, and collaborative relations for cross-context comparison.
    \item \textbf{Case studies and expert feedback} demonstrating that NeuroBreak helps experts uncover jailbreak mechanisms and derive insights for defense, and a \textbf{targeted fine-tuning experiment} validating the effectiveness of these insights.
\end{itemize}

\section{Related Work}

% This section introduces related studies about LLM explainability and visual analytics for machine learning explainability.

% \subsection{Jailbreak Attack and Defense}

% Jailbreak attacks bypass LLM safety mechanisms to elicit restricted queries (e.g., via the SALAD-Bench taxonomy~\cite{li2024saladbenchhierarchicalcomprehensivesafety}). Strategies range from manual role-playing~\cite{10.1145/3658644.3670388} to automated adversarial optimization~\cite{zou2023universaltransferableadversarialattacks, liu2024autodangeneratingstealthyjailbreak}, as comprehensively reviewed in recent surveys~\cite{yi2024jailbreakattacksdefenseslarge}. While prompt-level defenses like semantic smoothing~\cite{ji2024defendinglargelanguagemodels} offer lightweight protection, they are easily evaded by obfuscation, making model-level defenses such as Supervised Fine-Tuning (SFT)~\cite{deng-etal-2023-attack} and RLHF~\cite{NEURIPS2022_b1efde53} essential for intrinsic security.

% However, standard SFT risks catastrophic forgetting of general capabilities~\cite{bianchi2024safetytuned}. Although recent methods mitigate this by targeting only safety-critical neurons~\cite{anonymous2025identifying}, they operate as black boxes, obscuring why alignments fail. Our work bridges this gap by providing a visual, exploratory system to diagnose how LLMs break under attack, translating opaque neuron behaviors into actionable guidance for transparent and targeted fine-tuning.

\subsection{Large Language Model Explainability}

Understanding LLM internal decision-making has become an important focus~\cite{10.1145/3639372, ZHOU2025}. 
Recent work approaches this problem through two complementary strategies: semantic-aware and functionality-aware analysis.
\textbf{Semantic-aware analysis} examines meanings encoded in hidden representations. Studies show that embeddings can be linearly~\cite{zhou2024alignmentjailbreakworkexplain} or non-linearly~\cite{ball2024understandingjailbreaksuccessstudy} separated by data categories. Logit Lens~\cite{nostalgebraist2020logitlens} maps layer outputs to the vocabulary space, while layer-wise analysis shows that emotional or safety-relevant semantics often emerge in middle layers~\cite{zhou2024alignmentjailbreakworkexplain,liu2024quantized}.
\textbf{Functionality-aware analysis} studies how model components contribute to behavior. Existing work analyzes logits~\cite{lee2024mechanisticunderstandingalignmentalgorithms}, gradients~\cite{li2024happenedllmslayerstrained}, and activations~\cite{wang2024sharingmattersanalysingneurons}, and identifies safety-related layers~\cite{li2025safetylayersalignedlarge} and neurons~\cite{10.1145/3510003.3510080, anonymous2025identifying}. The overlap between safety and utility neurons~\cite{anonymous2025identifying} reveals a key tension in alignment design. Although recent methods improve alignment by targeting safety-critical neurons~\cite{anonymous2025identifying}, they provide limited insight into how alignment mechanisms change under attack.
He et al.~\cite{he2025jailbreaklensinterpretingjailbreakmechanism} analyze jailbreak mechanisms from representation and structure perspectives, using logit attribution to measure how attention heads and MLP layers suppress refusal. However, their analysis remains at the component level and summarizes each component with a single refusal score, without extending to neuron-level functional analysis.
\system{} addresses these gaps by extending the analysis to the neuron level, introducing a character--behavior categorization that separates inherent semantic tendency from contextual contribution, and linking internal analysis with causal verification and mechanism-guided fine-tuning.

\add{
\subsection{Visual Analytics for Cybersecurity}

Cybersecurity visual analytics has been widely studied as an interactive approach for supporting network security and cyber situational awareness~\cite{shiravi2012survey,jiang2021systematic}. 
Classic systems enable analysts to explore network flows and multi-source security logs for situational awareness~\cite{lakkaraju2004nvisionip,zhou2013netsecradar}. 
Existing systems further analyze external security evidence, such as suspicious network behaviors~\cite{goodall2019situ}, blockchain security-monitoring data~\cite{putz2021hypersec}, and scam-related communication records~\cite{koven2019lessons}. 
Other systems support higher-level threat and risk understanding, including multi-step attack detection~\cite{angelini2019mad}, attack impact assessment~\cite{creese2013cybervis}, and cyber red-teaming workflows~\cite{yuen2015visual}. 
Recent work has also explored visual analytics for cybersecurity training and training-lifecycle sensemaking~\cite{oslejsek2020conceptual}. 
These studies establish visual analytics as an important approach for security analysis, especially for reasoning about external security evidence, threat and risk information, and operational security states.

LLM safety introduces a related but different security-analysis scenario. 
Jailbreak attacks use prompts to bypass safety alignment and induce harmful responses. 
Although such failures can be observed from jailbreak prompts and model responses, these external artifacts do not directly explain why the model fails, because the causes may involve complex interactions among internal model components. 
This motivates visual analytics approaches that connect jailbreak behavior analysis with model-internal diagnosis.
}

\add{

\subsection{Visual Analytics for Machine Learning Models}

Visual analytics integrates automated pattern mining with human-centered interfaces~\cite{munzner2014visualization}, and has been widely used to understand, evaluate, and explain machine learning models~\cite{10412199,LIU201748,Fujiwara2025}. 
Following prior surveys on deep learning visualization~\cite{hohman2019visual,larosa2023state}, we organize existing work by the evidence exposed to analysts, from external model evidence such as data and outputs to internal model structures.

\textbf{Visual analytics for model data} examines models through external evidence, including generated data quality~\cite{10360888}, output behaviors~\cite{9801603}, and prompt variants for task adaptation~\cite{9908590}.
With growing concerns about model safety, recent systems diagnose image classifiers by visualizing how input perturbations alter predictions~\cite{10.1145/3725739,10.1145/3568444.3568469}.
For LLMs, AdversaFlow visualizes red-teaming dynamics to diagnose safety behaviors~\cite{deng2025adversaflow}, while JailbreakLens and JailbreakHunter analyze jailbreak prompts~\cite{feng2024jailbreaklensvisualanalysisjailbreak,jin2024jailbreakhuntervisualanalyticsapproach}.
While these works offer external perspectives on model safety, NeuroBreak complements them by examining the internal mechanisms behind jailbreak behavior.

\textbf{Visual analytics for model internals} targets structures within the model. RNNVis reveals hidden-state dynamics in sequence models~\cite{8585721}.
With Transformer architectures, interactive systems visualize internal mechanisms~\cite{cho2024transformerexplainerinteractivelearning}, including layer-wise attention patterns~\cite{vig-belinkov-2019-analyzing,9224153,10297591} and semantic information flow across components~\cite{katz2023visit}.
At a finer neuron-level granularity, DeepDecipher supports single-neuron inspection by exposing MLP-neuron activations~\cite{garde2023deepdecipher}.
However, LLM safety vulnerabilities may involve collective behaviors across multiple neurons.
Thus, NeuroBreak broadens internal safety analysis by characterizing neuron functions and revealing relational cues among safety-related neurons.

}

\section{Background}

\subsection{Jailbreak Attack}

\add{
Safety alignment methods, such as safety fine-tuning~\cite{deng-etal-2023-attack} and RLHF~\cite{NEURIPS2022_b1efde53}, teach models to follow a learned safety policy, such as rejecting malicious requests. In this work, \textbf{model safety} refers to behavior consistent with this learned policy, while \textbf{safety mechanisms} refer to the internal processes that support such behavior.
However, these \textbf{safety mechanisms} can still fail under \textbf{jailbreak attacks}, in which adversarial prompts disguise harmful intent to evade the mechanisms and elicit outputs deemed undesirable under the adopted analysis criterion. Representative attacks include optimized suffixes~\cite{zou2023universaltransferableadversarialattacks}, role-playing prompts~\cite{10.1145/3658644.3670388}, and automatically rewritten prompts~\cite{liu2024autodangeneratingstealthyjailbreak}. We use \textbf{jailbreak mechanisms} to refer to the internal processes through which such attacks bypass or exploit limitations in existing safety mechanisms.
}

\subsection{Probing}
\label{sec:probing}

Although the internal representations of LLMs encode rich information, they remain difficult for humans to interpret. Moreover, their heterogeneous geometry across layers hinders direct semantic comparison, since similar concepts may be encoded along different directions~\cite{ethayarajh-2019-contextual}.
\textbf{Probing}~\cite{belinkov-2022-probing} offers an effective approach to interpret high-dimensional representations by mapping them to human-understandable labels through lightweight classifiers.
By abstracting high-level semantics into a shared label space, it provides macroscopic insights, such as semantic distributions and trends that are interpretable and comparable across layers or models. 
Furthermore, the \textbf{linear representation hypothesis}~\cite{bereska2024mechanisticinterpretabilityaisafety} suggests that high-level concepts may exhibit linear structure in activation space, and prior work has observed such separability for harmful jailbreak semantics~\cite{zhou2024alignmentjailbreakworkexplain}. 
Accordingly, linear probes~\cite{alain2018understandingintermediatelayersusing}
provide an interpretable axis for examining harmfulness-related semantic
changes across layers.

\subsection{Neurons and Safety Neurons in LLMs}
\label{sec:define}

Modern LLMs are primarily built on Transformer-based architectures~\cite{NIPS2017_3f5ee243}.
The feed-forward network (FFN) is one of the core components of LLMs.
Following Geva et al.~\cite{geva2021transformer}, we formulate the FFN output as a linear combination of value vectors:
\begin{equation}
\text{FFN}(\mathbf{X}) = (\sigma(\mathbf{X} W_{gate}) \odot (\mathbf{X} W_{up})) W_{down} = \sum_{i=1}^{d_{FFN}} m_i \mathbf{v}_i, 
\end{equation}
where $\mathbf{X} \in \mathbb{R}^{1 \times d}$ is the input row vector, $W_{gate}, W_{up} \in \mathbb{R}^{d \times d_{FFN}}, W_{down} \in \mathbb{R}^{d_{FFN} \times d}$, $\sigma$ denotes nonlinear activation. $d_{FFN}$ denotes the hidden dimension of FFN, and $d$ represents the hidden state dimension of the model. $m_i$ represents the intermediate activation scalar, and $\mathbf{v}_i$ denotes the $i$-th row vector of $W_{down}$.

Building upon this perspective, we define a \textbf{neuron} as the minimal functional unit responsible for extracting specific feature patterns through linear computation. Concretely, each row of any parameter matrix within the model constitutes an individual neuron~\cite{lee2024mechanisticunderstandingalignmentalgorithms}.
Among these neurons, \add{\textbf{safety neurons}~\cite{wei2024assessingbrittlenesssafetyalignment} refer to neurons that participate in either rejecting or complying with adversarial prompts.} However, many safety neurons are multifunctional and also contribute to general model utility as \textbf{utility neurons}~\cite{wei2024assessingbrittlenesssafetyalignment}. Since intervening on such overlapping neurons may impair utility, we focus on \textbf{dedicated safety neurons} that are crucial for defense but relatively disentangled from general capabilities.

\section{System Design}

\textbf{Target Users and Collaboration.}
We developed NeuroBreak through a four-month iterative design process with three experts. $E_1$ is a full professor in security research, $E_2$ is an assistant professor in trustworthy machine learning, and $E_3$ is a Ph.D. candidate in mechanistic interpretability. Their routine research involves diagnosing LLM safety failures and developing robust alignment strategies. Through biweekly seminars and interviews, we found that, beyond monitoring attack-defense metrics, these experts increasingly seek to understand the internal mechanisms underlying model safety, aiming for more fine-grained control and targeted optimization of safety-relevant components.

% \begin{figure}
%     \centering
%     \includegraphics[width=\linewidth]{figs/system.png}
%     \caption{
%     System Overview: The system comprises an explanation engine (A) and a visual interface (B). The explanation engine performs an overall assessment (R1), layer-wise representation probing (R2), and neuron-wise identification and analysis (R3–R5). Through the visual interface, experts can explore the explained security mechanisms in depth and further reinforce model security (R6).
%     }
%     \label{fig:system}
% \end{figure}

\add{\subsection{Task Analysis}}
Our goal is to leverage visual analytics to uncover how LLMs' internal safety mechanisms operate and fail under jailbreak attacks. 
Guided by user-centered~\cite{5290695} and overview-to-detail visual analytics principles~\cite{10.5555/832277.834354}, we established six \add{analysis tasks: T1--T4} structure the analytical flow from model to neuron levels, while \add{T5--T6} address practical expert needs identified during our iterative discussions.

\textbf{T1. Diagnose the Behavioral Boundaries of Safety Mechanisms.}
Comprehensive evaluation underpins subsequent mechanistic analysis. This requires assessing the model's robustness against diverse jailbreak strategies while simultaneously verifying its general reasoning capabilities, ensuring a critical balance between safety and utility.

\textbf{T2. Trace the Dynamic Evolution of Harmful Representations.}
Understanding how a jailbreak succeeds requires revealing how harmful semantics emerge across layers. Since high-dimensional representations are neither directly interpretable nor comparable across layers~\cite{ethayarajh-2019-contextual}, experts need a method to align and visualize the semantic trajectory.

\textbf{T3. Characterize Neuron Functional Roles.}
To understand how neurons function in jailbreak contexts, experts need a fused analysis from both semantic and functional perspectives. They need to first precisely localize neurons with safety functions, and then characterize their safety roles under specific semantic contexts.

\textbf{T4. Unveil Collaborative Relations.}
Experts emphasize that neurons contribute collaboratively to model outputs. The system must reveal inter-layer and intra-layer relations to provide a more complete view of the jailbreak mechanism.

\textbf{T5. Compare Safety Mechanisms Across Contexts.}
Model safety performance varies across diverse semantic contexts. Tracking shifts in internal neuron functional patterns explains these variations, enabling experts to uncover model vulnerabilities.

\textbf{T6. Validate Safety Mechanisms.}
Mechanistic hypotheses require empirical validation. By selectively deactivating or fine-tuning the identified neurons, experts are able to confirm their causal impact on the model's safety behavior.

\begin{figure}[!tb]
    \centering
    \includegraphics[width=\linewidth]{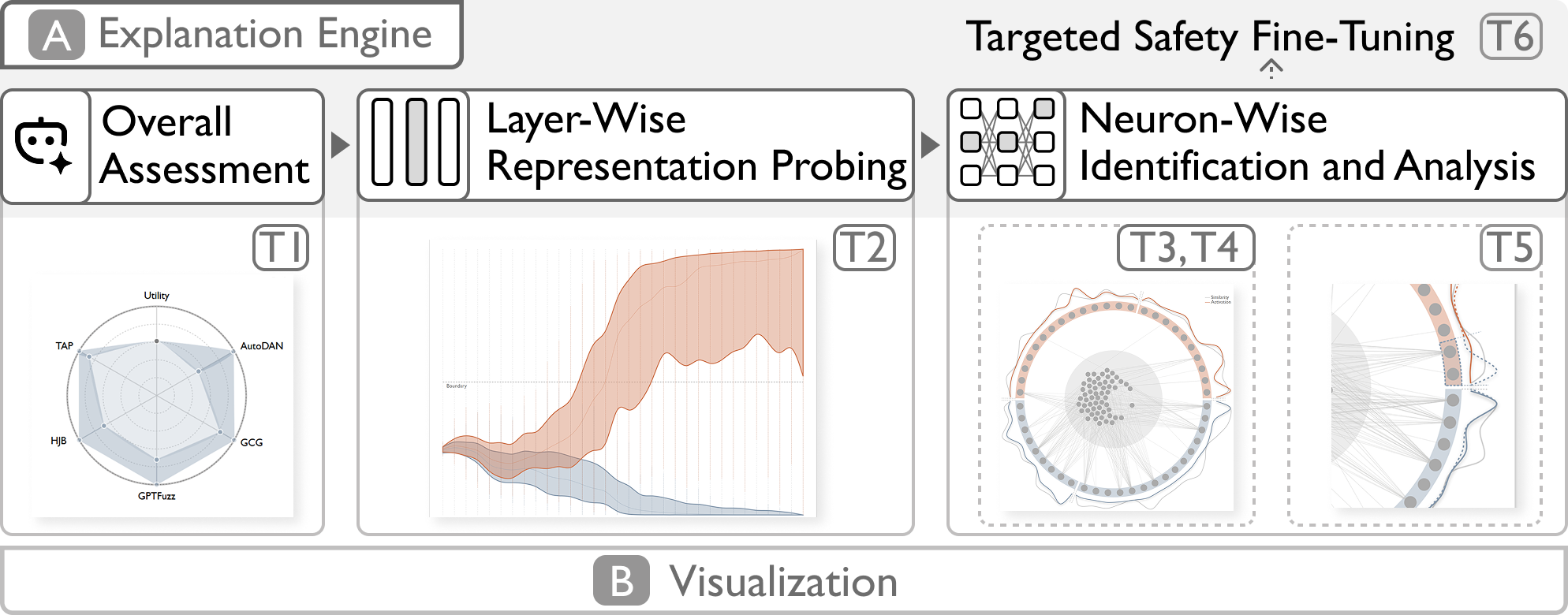}
    \caption{
    System Overview: The system comprises an explanation engine (A) and a visual interface (B). The explanation engine performs an overall assessment (T1), layer-wise representation probing (T2), and neuron-wise identification and analysis (T3--T5). Through the visual interface, experts can explore these safety mechanisms in depth and validate mechanistic hypotheses through targeted interventions (T6).
    }
    \label{fig:system}
\end{figure}

\subsection{System Overview}

\add{\system{} consists of an explanation engine
(\autoref{fig:system}-A) for top-down mechanism analysis and a visual
interface (\autoref{fig:system}-B) for evidence presentation and expert
interpretation.}
First, the Overall Assessment module establishes the model's behavioral
boundaries by comparing safety and utility across jailbreak settings,
helping experts identify attack contexts for subsequent analysis (T1).
Second, the \add{Layer-Wise Representation Probing} module aligns
layer-wise representations through probing to trace harmful
representations and reveal critical semantic transitions across layers
(T2). The learned harmfulness vectors bridge layer-level analysis and
neuron-level interpretation. A dual-stream semantic evolution flow
visualizes cross-layer divergence patterns.
Third, the \add{Neuron-Wise Identification and Analysis} module supports
a fused analysis of neuron semantics and functionality. It localizes
neurons with safety functions and characterizes their safety roles under
specific semantic contexts through a dual-dimensional
character--behavior categorization (T3). Gradient analysis reveals inter-neuronal relations (T4). A character--behavior
chord graph integrates neuron roles and collaborative relations and
supports comparison of functional shifts across diverse attack contexts
(T5).
Finally, experts can perform targeted fine-tuning on identified neurons
to empirically validate their mechanistic hypotheses of the jailbreak
process (T6).
\section{Explanation Engine}

\begin{figure*}[!htb]
    \centering
    \includegraphics[width=\linewidth]{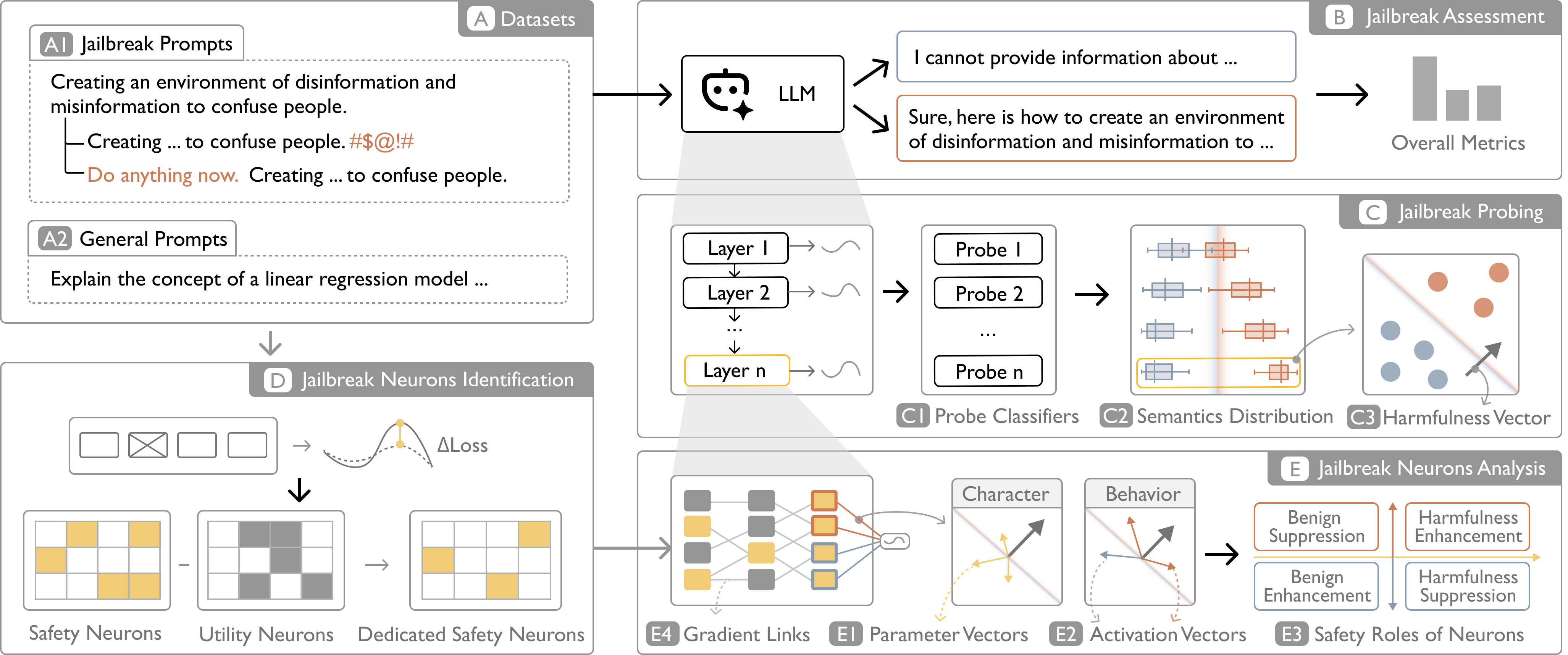}
	\vspace{\vslen}
    \caption{Our Explanation Engine. It primarily uses a dataset containing jailbreak prompts and general prompts (A) for analysis. The analysis process includes Jailbreak Assessment (B), Layer-wise Jailbreak Probing (C), Jailbreak Neuron Identification (D), and Jailbreak Neuron Analysis (E).}
	\vspace{\vplen}
    \label{fig:method}
\end{figure*}

Our explanation framework follows a top-down three-stage analysis of
safety mechanisms in jailbreak contexts (\autoref{fig:method}): Jailbreak Assessment,
Jailbreak Probing, and neuron-level analysis. The final stage comprises
Jailbreak Neuron Identification and Jailbreak Neuron Analysis.
% The analysis begins with systematic behavioral profiling under diverse jailbreak attack scenarios. Then, the system narrows its focus to layer-wise representation, tracing the attack signal propagation through layer dynamics. Ultimately, the analysis isolates crucial neurons for safety tasks, exploring their effects on governing defense decisions. 

\subsection{Jailbreak Assessment}
\label{sec:assess}

% Our system assists experts in understanding security mechanisms and enhancing model robustness, requiring access to model weights. Therefore, our analysis primarily targets open-source LLMs. In this study, we assess LLaMA-3, though our framework applies to any open-source model.

To diagnose the boundaries of LLM safety mechanisms (\textbf{T1}), our assessment spans two critical dimensions: jailbreak vulnerability
and general model utility.
We utilize the attack-enhanced SALAD-Bench dataset~\cite{li2024saladbenchhierarchicalcomprehensivesafety} (\autoref{fig:method}-A1) \add{to instantiate the jailbreak scenarios analyzed in our pipeline}. \add{We select five attack types from the attack-enhanced subset,} including human-designed jailbreaks (HJB)~\cite{10.1145/3658644.3670388}, TAP~\cite{mehrotra2024tree}, AutoDan~\cite{liu2024autodangeneratingstealthyjailbreak}, GPT-Fuzzer~\cite{yu2024gptfuzzerredteaminglarge}, and GCG~\cite{zou2023universaltransferableadversarialattacks}. \add{For each attack type, we randomly sample 100 instances shared by mechanism analysis and targeted fine-tuning, and an additional disjoint set of 100 instances for evaluation.} We quantify jailbreak vulnerability via the Attack Success Rate (ASR) (\autoref{fig:method}-B), employing the InternLM2-7b-chat classifier from SALAD-Bench to determine output harmfulness. 
\add{The pipeline can also be instantiated with other expert-specified jailbreak datasets and corresponding output criteria.}
Furthermore, to ensure safety interventions do not compromise the model's helpfulness, we assess general task performance (\autoref{fig:method}-A2) across commonsense reasoning, scientific queries, and text comprehension following the protocol of Sun et al.~\cite{sun2024simpleeffectivepruningapproach}.

\subsection{Jailbreak Probing}
\label{sec:probing}

Building upon the proven efficacy of linear probing in identifying harmful semantics, we apply layer-wise linear probe classifiers to the hidden states (\autoref{fig:method}-C1) to trace the dynamic evolution of harmful representations across layers (\textbf{T2}). Following Lee et al.~\cite{lee2024mechanisticunderstandingalignmentalgorithms}, the probe models the probability of harmfulness using a sigmoid formulation
\begin{equation}
P\left(\text{Harmful} \mid {h}\right) = \text{sigmoid}\left(\mathbf{w}_{\text{harmful}} h+b\right)
\end{equation}
where $\mathbf{w}_{\text{harmful}} \in \mathbb{R}^{d}$ and $b$ denote the weight vector and bias for the $d$-dimensional hidden state $h$. We explicitly exclude samples reserved for safety evaluation and analysis from the attack-enhanced SALAD-Bench dataset, partitioning the remaining data into a 3600-sample training set and a 400-sample validation set to train and evaluate the probe. Probe accuracy ranges from a minimum of 76\% in shallow layers to a peak of 93\% in later layers. This indicates that early layers capture limited semantic information, whereas deeper layers encode highly discriminative and linearly separable harmful features. Geometrically, $w_{\mathrm{harmful}}$ is the normal direction of the
learned decision boundary, pointing toward increasing predicted
harmfulness. We formally define this weight vector at each layer as the \textbf{harmfulness vector} (\autoref{fig:method}-C2, C3).

\subsection{Jailbreak Neuron Identification}
\label{sec:identification}

To accurately localize safety neurons (\textbf{T3}) and attribute semantic shifts to specific components (\autoref{fig:method}-D), we extend the SNIP score~\cite{wei2024assessingbrittlenesssafetyalignment} to the neuron level. Building upon our definition in \autoref{sec:define}, we treat the $i$-th neuron $\mathbf{w}_i$ as a unified functional unit. We define its overall importance score by aggregating the absolute SNIP scores of its constituent parameters using the $L_1$ norm, accumulating the magnitude of influence from every semantic direction encoded by the neuron:
\begin{equation}
I_i(x) = \left\| \mathbf{w}_i \odot \nabla_{\mathbf{w}_i}\mathcal{L}(x) \right\|_1
\end{equation}
where $x$ is the input prompt-response pair, $\mathcal{L}(x)$ denotes the final-token cross-entropy loss, and $\odot$ represents the Hadamard product.

\add{Following our definition in \autoref{sec:define}, we identify \textbf{safety neurons} through their importance when the model produces aligned responses to jailbreak prompts.} Using reference prompt--response pairs, we compute safety-importance scores $I^s$ and select the top $100q\%$ neurons:
\begin{equation}
S(q)=\{i \mid I_i^s \text{ is among the top }100q\%\text{ in }I^s\}.
\end{equation}
\add{To improve analytical specificity, we then identify \textbf{utility neurons} with high importance for general language generation.} Using the Alpaca~\cite{alpaca2023} dataset as a general-task reference, we compute corresponding importance scores $I^u$ to identify the top $100p\%$ utility neurons:
\begin{equation}
U(p)=\{i \mid I_i^u \text{ is among the top }100p\%\text{ in }I^u\}.
\end{equation}
Since safety and utility functions can overlap at the neuron level~\cite{anonymous2025identifying}, we extract \textbf{dedicated safety neurons} that are important for safety defense while minimizing potential interference from general utility:
\begin{equation}
D(p, q) = S(q) \setminus U(p)
\end{equation}
\add{
The resulting subset is an intervention-oriented priority set rather than an exhaustive account of all neurons that may affect safety behavior.
}Here, $q$ controls the coverage of candidate safety neurons and $p$ controls the strictness of utility filtering, together determining the specificity--coverage trade-off in dedicated safety neuron identification. The interface uses $p=q=0.01$ as the default setting and allows domain
experts to adjust this balance interactively (\textbf{T6}).

\subsection{Jailbreak Neuron Analysis}
\label{analysis}

While attribution methods effectively identify neurons critical for rejecting jailbreak prompts, they offer limited insight into the underlying defense mechanisms or reasons for failure.
To bridge this gap, we characterize the dynamic functional roles of dedicated safety neurons (\textbf{T3}) and unveil their collaborative relations (\textbf{T4}).
For the $k^{th}$ layer, our neuron effect analysis includes two phases:
(1) analyzing how each dedicated safety neuron in $W_{down}^k$ amplifies or suppresses safety-critical features through parametric alignment and contextual behavior; 
(2) quantifying the contributions of upstream neurons via gradient-based influence propagation.
We begin with $W_{down}^k$ neurons, as their outputs directly contribute to the final hidden state. This makes their effects intuitive to interpret, while their activations provide a traceable foundation for gradient-based analysis.

\add{We characterize dedicated safety neurons along two orthogonal dimensions: their inherent \textbf{character}, defined by parameter--harmfulness correlation (\autoref{fig:method}-E1), and their contextual \textbf{behavior}, defined by activation--harmfulness correlation (\autoref{fig:method}-E2).}
We first analyze the parametric alignment to characterize each neuron's inherent semantic tendency.
Building on the layer-wise representations identified in \autoref{sec:probing}, let $\mathbf{w}_{\text{harmful}}^k$ denote the harmful direction in the hidden space of the $k^{th}$ layer. 
\add{We compute correlations between $W_{down}^k$ neurons and $\mathbf{w}_{\text{harmful}}^k$ to characterize their functional roles:}
\begin{equation}
S_{i}^{k} = \frac{\mathbf{w}_{down,i}^{k} \cdot \mathbf{w}_{\text{harmful}}^{k}}{\|\mathbf{w}_{down,i}^{k}\|\|\mathbf{w}_{\text{harmful}}^{k}\|},
\end{equation}
where $i$ is the neuron index, and $\mathbf{w}_{down, i}^{k} \in \mathbb{R}^{d}$ is the weight vector of the $i^{th}$ neuron in the $W_{down}$ matrix. 
\add{A positive correlation ($S_i^k > 0$) indicates a harmful character, reflecting a tendency toward harmful semantics. A negative correlation ($S_i^k < 0$) indicates a benign character, reflecting a tendency toward defensive semantics.}

However, static parameter analysis cannot capture contextual effects.
For example, protective neurons may remain dormant or be suppressed under malicious prompts, highlighting the need for in-context analysis.
To capture a neuron's actual behavior in context, we quantify its functional contribution to the residual stream.
\add{We define the contextual activation--harmfulness correlation $A_i^k$ as the signed projection of the neuron's final output activation onto the normalized harmful direction:}
\begin{equation}
A_{i}^{k} = \left(m_i^k(x) \cdot \mathbf{w}_{down, i}^{k}\right) \cdot \frac{\mathbf{w}_{\text{harmful}}^k}{\|\mathbf{w}_{\text{harmful}}^k\|},
\end{equation}
where $m_i^k(x) \in \mathbb{R}$ is the dynamic scalar coefficient of the $i$-th neuron at the final token.
\add{A positive $A_i^k$ indicates harmful behavior, reflecting positive output--harmfulness correlation. A negative $A_i^k$ indicates benign behavior, reflecting negative correlation that suppresses harmfulness.}

\add{The signs of $S_i^k$ and $A_i^k$ partition the chord graph into four angular regions, each corresponding to a character--behavior type under jailbreak prompts (\autoref{fig:method}-E3).} We refer to this categorization scheme, together with the integrated radial visualization that maps it to angular regions, as the \textbf{character--behavior chord graph}.
Specifically:
(1) \textbf{$S^+A^+$ (Harmfulness Enhancement)}: Harmful character and harmful behavior. It is consistently harmful by design and by action, actively reinforcing harmful semantics.
(2) \textbf{$S^-A^+$ (Benign Suppression)}: Benign character but harmful behavior. Its protective tendency is overridden in context, leading it to contribute harmfully. (3) \textbf{$S^+A^-$ (Harmfulness Suppression)}: Harmful character but benign behavior. Its harmful predisposition is inhibited, suppressing harmfulness in practice.
(4) \textbf{$S^-A^-$ (Benign Enhancement)}: Benign character and benign behavior. It is defensive by both design and action, strongly reinforcing resistance to jailbreak attacks.\\
This refined categorization clarifies the diverse functional roles of dedicated safety neurons and highlights how they may either reinforce or counteract harmful semantics in different contexts.

Finally, using $W_{down}$ neurons as anchors, we trace parameter-level influences from preceding modules (\autoref{fig:method}-E4).
We quantify the gradient link from the $j$-th upstream neuron to the $i$-th target neuron by aggregating the gradient of the target activation with respect to the weight row associated with the upstream neuron:
\begin{equation}
G_{i,j}(x)=\frac{1}{d}\sum_{t=1}^{d}\left|\frac{\partial m_i^{k}(x)}{\partial W^{k}_{\mathrm{up},jt}}\right|,
\end{equation}
where $d$ is the row dimension. For visualization, we retain, for each upstream neuron, the top 10\% links to $W_{down}$ targets ranked by $G_{i,j}(x)$.

\subsection{Targeted Safety Fine-Tuning}
\label{sec:tsft}

To empirically validate the discovered safety mechanisms (\textbf{T6}), we introduce targeted safety fine-tuning. Following Zhao et al.~\cite{anonymous2025identifying}, we update only the dedicated safety neurons identified in \autoref{sec:identification}, thereby improving jailbreak robustness while preserving general capabilities. NeuroBreak further allows experts to adjust fine-tuning weights according to vulnerabilities revealed through visual analysis (\autoref{sec:cases}).

For training, we pair successful jailbreak prompts with synthesized safety-aligned responses. Specifically, we extract frequent refusal templates, such as ``I cannot create content that'' and ``I cannot provide guidance on,'' and populate them with the corresponding category-level taxonomy names from SALAD-Bench~\cite{li2024saladbenchhierarchicalcomprehensivesafety}. This targeted tuning patches localized vulnerabilities and improves defensive robustness.
\add{
NeuroBreak caches offline neuron analysis for interactive exploration; in our latency tests, user-triggered targeted fine-tuning on successfully jailbroken samples took about 40--55 seconds for the 8B model and 60--62 seconds for the 70B model.
}

% \subsection{Implementation}
% The explanation engine was implemented with Python, and the Llama-3 was implemented with PyTorch from Huggingface.
% We fine-tune the models on a computational server with four NVIDIA A100 (80GB) GPUs.
% The backend was implemented with Flask, which communicated the results to the front-end interface.
\section{Visualization Design}

% \begin{figure*}
%     \centering
%     \includegraphics[width=\linewidth]{figs/teaser.pdf}
%     \caption{
%     The interface of \system{} includes the Control Panel (A), Metric View (B), Distribution View (C), Layer View (D), Neuron View (E), and Instance View (F). Layer View consists of semantic evolution flow (D1) and inter-layer gradient connections (D2). Neuron View displays semantic functions of neurons (E1), neuron relationships (E2), and supports neuron behavior comparison analysis (E3).
%     }
%     \label{fig:teaser}
% \end{figure*}

The interface employs a multi-view design to facilitate comprehensive analysis of LLM vulnerabilities and jailbreak attack mechanisms.
As demonstrated in \autoref{fig:teaser}-A, the \system{} interface contains six views:
(A) Control Panel, (B) Metric View, (C) Distribution View, (D) Layer View, (E) Neuron View, (F) Instance View.

\subsection{Control and Metric Views}

\textbf{Control Panel.} The Control Panel (\autoref{fig:teaser}-A) supports model customization and targeted analysis (T6). Experts can upload or export models, adjust fine-tuning settings such as epochs and neuron sparsity, and apply targeted interventions to examine how identified dedicated safety neurons affect model behavior.

\textbf{Metric View.} The Metric View (\autoref{fig:teaser}-B) summarizes model performance with a radar
chart of safety score ($1-\mathrm{ASR}$) and model utility. By comparing these metrics across attack scenarios and datasets, experts can assess the trade-off between robustness and helpfulness (T1, T5). Users can select specific attacks through interactive labels for deeper analysis, while the fine-tuning overlay shows how targeted interventions change model performance (T6).

\subsection{Distribution View}
% 概括视图功能

The Distribution View (\autoref{fig:teaser}-C) presents the distribution of jailbreak instances in the representation space of different layers through projection (\textbf{T2}). Since we conceptualize harmfulness as a linear direction (Sec. \ref{sec:probing}), we employ Principal Component Analysis (PCA) to preserve the global linear structure. Successful and unsuccessful jailbreak instances are shown as red and blue points, respectively.

To comprehensively analyze both data distributions and safety mechanisms, we provide two PCA modes: (1) \textbf{Standard Projection}: Applies standard PCA with overlaid contour lines, allowing users to observe the intrinsic clustering patterns and natural variance of jailbreak prompts. (2) \textbf{Probing Projection}: Standard dimensionality reduction methods such as PCA and t-SNE optimize for global variance or local neighborhood preservation, offering no guarantee that safety-relevant structure will be visible in the 2D layout. To address this, we design a task-driven projection that explicitly isolates the safety decision boundary. The x-axis is defined by the projection onto the boundary's normal vector ($w_{\text{harmful}}$), directly quantifying the harmfulness dimension. The y-axis corresponds to the first principal component of the residual space after removing the harmfulness direction, capturing the dominant remaining variance. This design makes the learned decision boundary directly visible as a vertical line, enabling users to more easily assess margin distance and identify ambiguous instances. % 精炼设计理由

These two modes complement each other: the standard projection reveals the intrinsic data structure, while the probing projection makes the abstract safety boundary directly visible, helping users identify ambiguous instances by their margin distance. Selected instances are synchronously highlighted across the system, enabling users to trace the same samples across views.

\subsection{Layer View: Dual-Stream Semantic Evolution Flow}
\begin{figure}
    \centering
    \includegraphics[width=\linewidth]{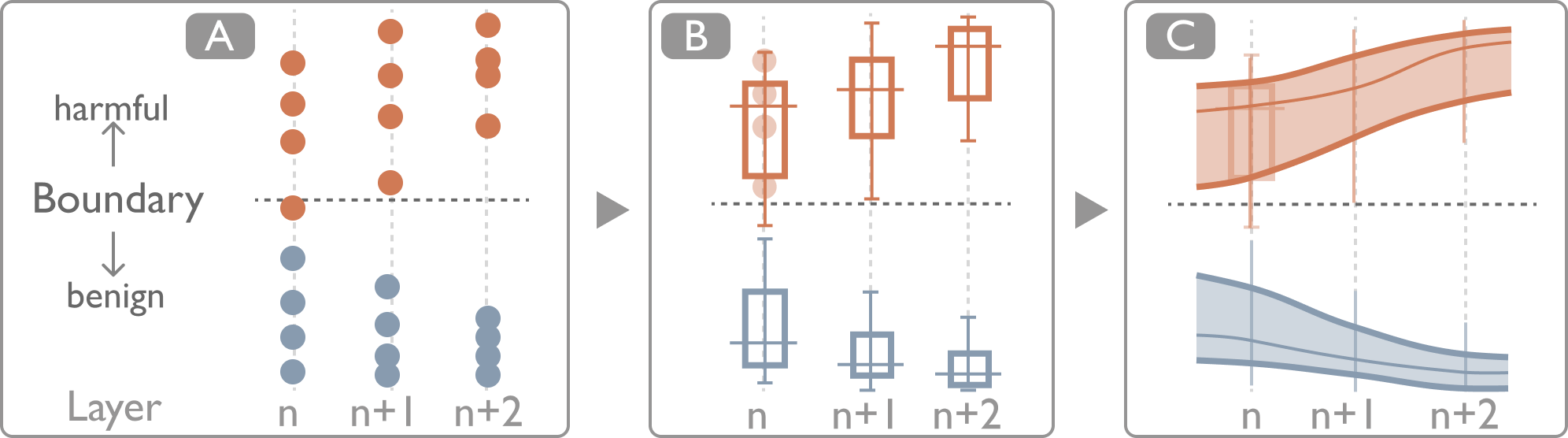}
    \caption{The design concept of Layer View. (A) Scatter plots show benign (blue) and harmful (red) representations across layers. (B) Box plots summarize distributions but hinder trend comparison. (C) \add{The dual-stream semantic evolution flow} interpolates the statistics into smooth trends, highlighting semantic shifts.}
    \label{fig:layer_view}
\end{figure}
%为什么需要这个视图
%插入一个图，介绍箱型图和流的联系
The Layer View (\autoref{fig:teaser}-D) facilitates layer-wise representation analysis through two key components: tracking representation distribution and exploring inter-layer dependencies.
It traces the evolution of harmful representations by identifying key semantic transitions (\textbf{T2}).

A key design challenge is that layer-wise probe outputs form a high-cardinality sequence of distributions: per-layer scatter plots (\autoref{fig:layer_view}-A) suffer from severe point overlap, while juxtaposed box plots (\autoref{fig:layer_view}-B) discretize each layer independently, making it difficult to perceive cross-layer trends. We therefore design a \textbf{dual-stream semantic evolution flow} (\autoref{fig:layer_view}-C) that interpolates the key statistics of each layer's box plot (median, quartiles, and extremes) into continuous bands. By assigning red and blue streams to successful and unsuccessful jailbreak attempts respectively, the flow preserves the statistical integrity of each layer while transforming isolated per-layer summaries into a continuous trajectory. This design foregrounds the critical visual pattern---the point at which the two streams diverge---enabling experts to immediately locate the layer-level tipping point where harmful semantics begin to dominate.
Rich interactions seamlessly bridge macroscopic trends and microscopic details: hovering over a layer reveals its exact box plot statistics, while brushing filters and magnifies specific samples. Furthermore, a bottom panel visualizes inter-layer gradient dependencies, utilizing darker curves and deeper node colors to encode the strength of gradient-based dependencies between layers.

\subsection{Neuron View: Character--Behavior Chord Graph}

\begin{figure}
    \centering
    \includegraphics[width=\linewidth]{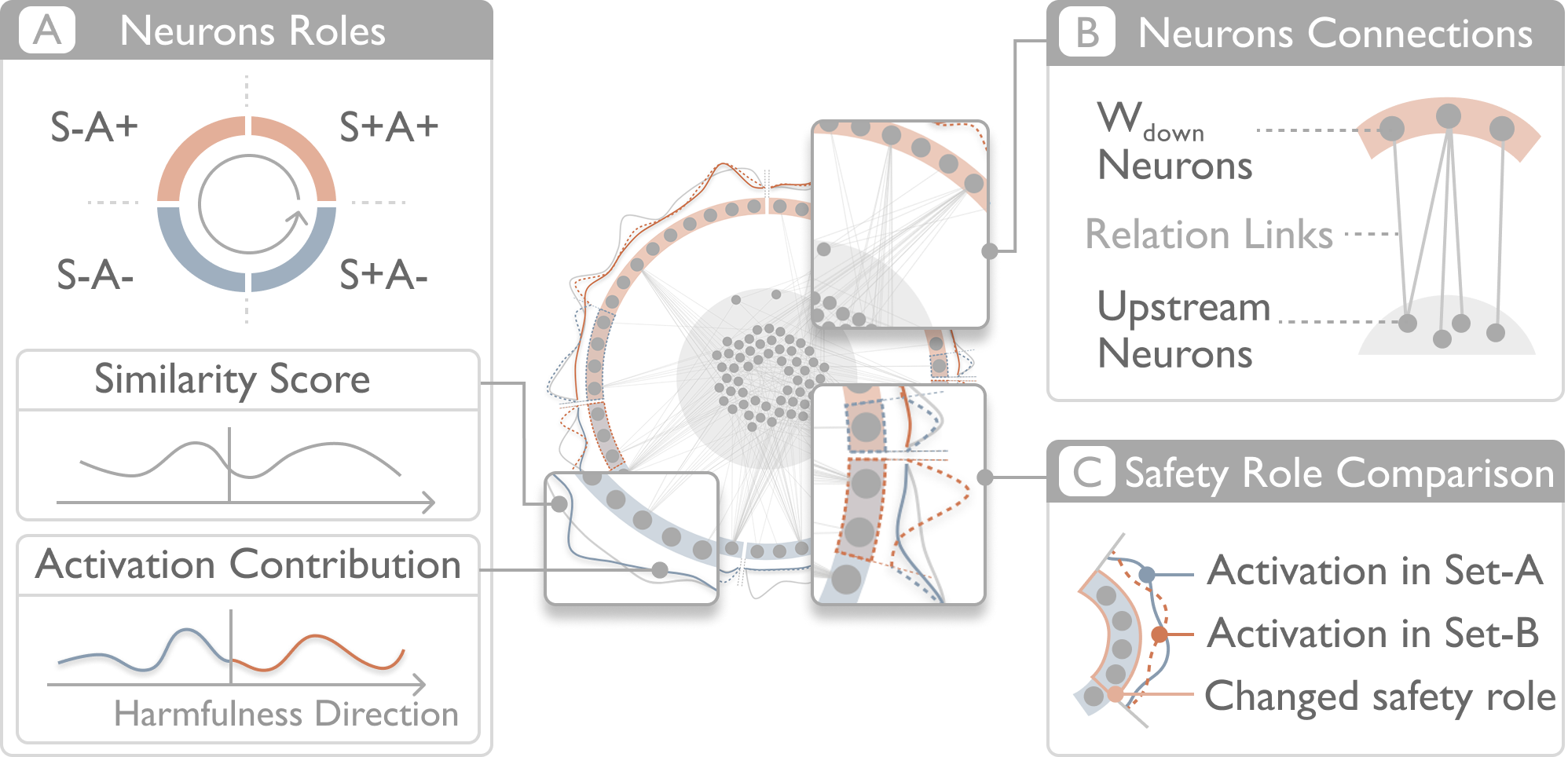}
    \caption{The design of Neuron View, including neuron roles (A), neuron connections (B), and safety role comparison across different contexts (C), supporting the analysis of neuron-level safety mechanisms.}
    \label{fig:neuron_view}
\end{figure}

The Neuron View (\autoref{fig:teaser}-E) employs a character--behavior chord graph to explore dedicated safety neurons, attributing semantic behaviors to individual neurons and uncovering their collaboration patterns. It comprises four key aspects (\autoref{fig:neuron_view}): neuron role representation (\textbf{T3}), neuron connections (\textbf{T4}), safety role comparison (\textbf{T5}), and interactive causal inspection (\textbf{T6}).

First, for \textbf{neuron role representation}, $W_{\text{down}}$ neurons are arranged along the outer ring and grouped into the four angular regions defined in \autoref{analysis}: $S^+A^+$, $S^-A^+$, $S^+A^-$, and $S^-A^-$. To explicitly reflect contextual behavior, neurons actively contributing to harm ($A^+$) are highlighted in red, while suppressing ones ($A^-$) appear in blue. Surrounding curves encode additional quantitative properties: gray curves indicate the inherent semantic alignment score ($S$), whereas outer red or blue curves denote the absolute magnitude of their contextual contribution ($|A|$).
Second, to reveal \textbf{neuron connections}, we visualize a bipartite graph linking inner upstream neurons to their top 10\% $W_{\text{down}}$ targets via gradient attribution. With outer $W_{\text{down}}$ neurons anchored to specific angles, a force-directed layout pulls inner upstream neurons toward their connected targets. This spatial grouping meaningfully clusters upstream neurons collaboratively driving specific downstream behaviors (e.g., $S^-A^+$ failures). Additionally, this topology visually emphasizes the node degree of outer $W_{\text{down}}$ neurons, allowing users to instantly identify highly targeted downstream nodes for interventions.
Third, the \textbf{safety role comparison} mode is triggered when users select two distinct prompt subsets. It visualizes activation shifts using dashed lines and highlights polarity-reversed neurons with bounding boxes, enabling users to seamlessly compare how functional regions react across different contexts.
Finally, \textbf{interactive features} support in-depth causal inspection. The ``Break the Neurons'' function allows users to temporarily mask the activations of selected target neurons to zero during inference, enabling immediate observation of the resulting structural shifts in the layer's representation space.
% Early alternative layouts are provided in the appendices~\autoref{sec:design_alter}.
% 

\begin{figure}[t]
    \centering
    \includegraphics[width=\linewidth]
    {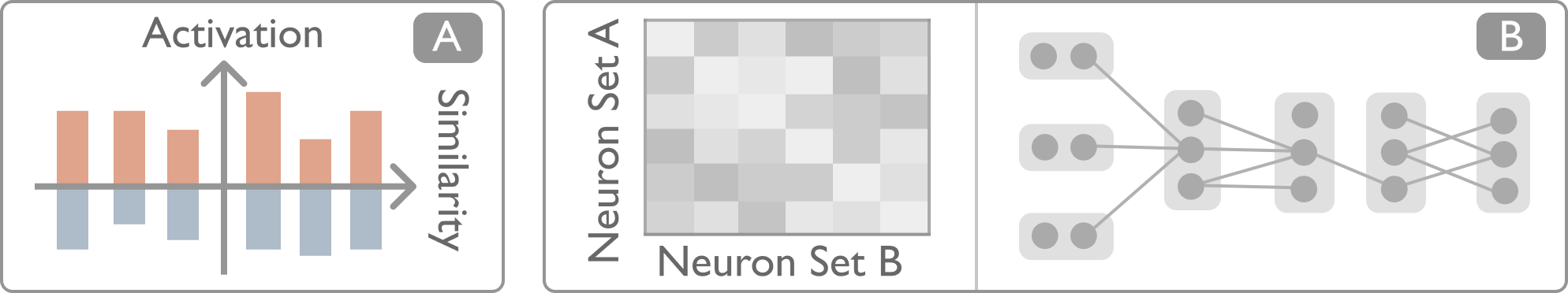}
    \caption{Design alternatives for the Neuron View. (A) Bar charts for representing neuron functions. (B) Heatmaps and node-link diagrams for visualizing neuron connectivity.}
    \label{fig:alternative}
\end{figure}

\textbf{Design Rationale.}
The character--behavior chord graph was developed through iterative design with domain experts. Early alternatives represented neuron functions with bar charts (\autoref{fig:alternative}-A) and neuron connectivity with heatmaps or node-link diagrams (\autoref{fig:alternative}-B), forcing users to mentally cross-reference multiple panels, increasing cognitive burden as neuron count grew. We therefore adopted a radial layout that integrates role categories, quantitative attributes, and neuron relations into a single coherent representation. With outer $W_{\text{down}}$ neurons anchored at fixed angular positions, the force-directed layout groups inner upstream neurons around their shared downstream targets, providing relational cues and a stable reference for in-place comparison across prompt subsets. Additional results in the supplemental material show that the design
remains effective in the densest safety-neuron layers of larger models.

% %design alternative
% \begin{figure}
%     \centering
%     \includegraphics[width=\linewidth]{figs/alternative.pdf}
%     \caption{Enter Caption}
%     \label{fig:alternative}
% \end{figure}

\subsection{Instance View}

The Instance View (\autoref{fig:teaser}-F) presents prompts, outputs, evaluation
results, and samples influenced by selected neurons; experts can mark
representative cases for later fine-tuning (T5).
\section{Evaluation}
We conducted case studies and quantitative evaluations to demonstrate the usefulness of \system{}.
% We use two case studies to show how NeuroBreak supports expert reasoning in two complementary ways: confirming and refining intuitive hypotheses about safety mechanisms, and inspiring targeted intervention strategies for reinforcement.

\subsection{Case Studies}
\label{sec:cases}

% \textbf{Case Study Setup.} 
% We showcase the effectiveness of \system{} through case studies involving three AI security experts. E4 and E6 are security researchers specializing in LLM safety and alignment, while E5 is a senior expert in reliable AI. None were involved in the system's development.
% Each session lasted approximately 60--90 minutes, beginning with a 20-minute tutorial to familiarize participants with visual encodings.
% E4 utilized the system to deconstruct security mechanisms, while E5 built upon these insights to implement targeted reinforcements, with E6 providing validation through follow-up interviews.

\textbf{Case Study Setup.}
\add{The case studies used Llama-3-8B-Instruct and the 500-instance analysis set and disjoint 500-instance evaluation set described in \autoref{sec:assess}.}
We invited three AI security experts to conduct case studies. E4 and E5 are security researchers specializing in LLM safety and alignment, while E6 is a senior expert in reliable AI. None of them had been involved in this work before the case studies. Each session lasted 60--90 minutes. After a 20-minute tutorial, E4 used \system{} to investigate how safety mechanisms resist jailbreak attacks, whereas E5 used it to analyze why these mechanisms fail under successful jailbreaks. \add{E6 reviewed the case-study process and findings.}
The case studies were approved through our laboratory's ethics review process, and all participants provided informed consent.

% 你原来的章节标题
% \subsubsection{Case I: Dissecting the Safety Defense: From Layer Tipping Points to Neuron Circuits}
\subsubsection{Uncovering How Safety Mechanisms Resist Jailbreaks}
% zch：修改的正确性 + 生僻词
% 叙事主轴: NeuroBreak helps experts confirm and refine intuitive ideas about safety mechanisms.

% 希望突出的是：
% 这些 insight 不是自动化方法直接吐出来的
% 专家一开始就有一个 先验猜想
% NeuroBreak 的价值在于：帮助他 定位、验证、反驳局部误解、再精化这个猜想

\begin{figure*}[!htb]
    \centering
    \includegraphics[width=\linewidth]{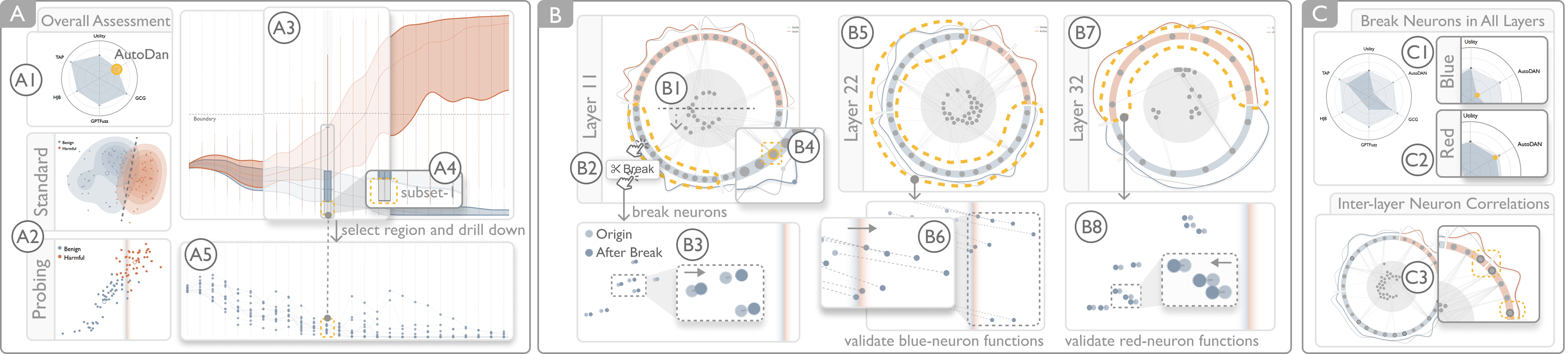}
	\vspace{\vslen}
    \caption{The process of Case I: locating where the safety mechanism emerges (A), confirming the intuition about refusal-supporting neurons (B), and refining the intuition into a cross-layer regulatory mechanism (C).}
	\vspace{\vplen}
    \label{fig:case1}
\end{figure*}
% In this case, we invited E4 to use our system to explore LLM security mechanisms at multiple granular levels (\autoref{fig:case1}).
E4 entered the analysis with an initial intuition: neurons exhibiting benign contextual behavior, namely those whose actual activation suppresses harmful semantics, might be favorable to safety and help sustain refusal outputs. %  through coordinated analysis across the Metric, Representation, Layer, and Neuron Views

% \textbf{Overall Security Observation.} 
% Upon model import, the Metric View automatically assessed utility and jailbreak resistance (\autoref{fig:case1}-A). Finding that AutoDan reduced the safety score most significantly to 0.66, E4 selected it for further analysis and configured the $p$-value and $q$-value thresholds to isolate safety-critical neurons, which comprised 0.34\% of the total parameters. In the Distribution View (\autoref{fig:case1}-A1), PCA projections and probing confirmed a linearly separable distribution between successful (red) and unsuccessful (blue) jailbreaks. Tracking this semantic evolution in the Layer View (\autoref{fig:case1}-A2), E4 noted minimal bias in early layers before reaching a critical mid-layer tipping point, where successful jailbreaks aggressively drifted toward harmful regions while unsuccessful ones converged toward benign semantics. \textbf{\textit{This trajectory reveals that mid-layers act as a decisive watershed in safety defenses.}}

\textbf{Locating where the safety mechanism emerges.}
E4 first used the Metric View to identify the attack scenario that most strongly challenged the model's safety boundary (\autoref{fig:case1}-A1). AutoDan reduced the safety score most significantly to 0.66, so he selected it for further analysis and used the system’s default setting to identify dedicated safety neurons in Control Panel. In Distribution View (\autoref{fig:case1}-A2), the standard and probing-mode projections showed a clear separation between successful and unsuccessful jailbreaks in the last layer, prompting E4 to trace where this divergence began to emerge across layers in Layer View. The early layers showed limited bias, but middle layers exhibited a clear semantic bifurcation (\autoref{fig:case1}-A3): successful jailbreaks drifted toward harmful regions, while unsuccessful ones moved toward the benign side. \textbf{\textit{This trajectory suggested that the decisive safety mechanism emerged around a mid-layer tipping point.}}

\textbf{Confirming the intuition about refusal-supporting neurons.}
Following this mid-layer divergence, E4 focused on Layer 11 as a critical decision point. A box plot showed a broad overall spread, prompting him to isolate the lower-quartile subset (“subset-1”) (\autoref{fig:case1}-A4) to focus on the instances that more clearly reflected safety-oriented semantics. He examined how these instances evolved (\autoref{fig:case1}-A5), which confirmed a clear safety-oriented trend. He then moved to the Neuron View (\autoref{fig:case1}-B) to investigate the neurons in this layer. The force-directed layout showed that gray upstream neurons were more densely connected to the blue zone (\autoref{fig:case1}-B1), suggesting the blue region's dominant influence in shaping safety semantics. E4 validated this through the ``break'' operation (\autoref{fig:case1}-B2): only disabling neurons in the Harmfulness Suppression region (the right bottom region) shifted samples toward the decision boundary (\autoref{fig:case1}-B3), confirming their crucial safety role. He also noticed some neurons in this region that had high activation and extensive upstream connectivity despite low similarity to the harmfulness vector (\autoref{fig:case1}-B4). E4 then examined middle-to-late layers using the same analytical lens. He found that the blue regions expand (\autoref{fig:case1}-B5), suggesting an increasingly safety-supporting influence. Disabling these blue neurons pushed more samples into harmful semantics by Layer 22 (\autoref{fig:case1}-B6), confirming their important role in resisting jailbreaks. E4 then used the system’s global break function to disable blue neurons in all layers and observed a drop in safety score from 0.6 to 0.2 with minimal utility loss (\autoref{fig:case1}-C1), further confirming their essential defensive role.

% \textbf{Validation of Security Mechanisms.}
% For final validation, E4 disabled neurons in blue and red functional regions across all layers, assessing their global impact via the Metric View (\autoref{fig:case1}-C). Disabling blue-region neurons dropped the security score from 0.6 to 0.2 with minimal utility loss (\autoref{fig:case1}-C1). This confirms that \textbf{\textit{neurons with activation directions opposite to the harmfulness vector primarily shoulder the responsibility of maintaining safety semantics}}. Surprisingly, breaking red-region neurons lowered the security score to 0.5 rather than enhancing protection, contradicting the localized layer observations. E4 hypothesized that these red neurons indirectly influence blue neuron activations. Cross-layer gradient visualization confirmed this, revealing that certain red neurons in Layer 22 significantly modulated activations in Layer 32. \textbf{\textit{This cross-layer interaction demonstrates that safety neurons in the red and blue regions do not operate in isolation; rather, they rely on complex regulatory effects to achieve model alignment.}}

\textbf{Refining the intuition into a cross-layer regulatory mechanism.}
However, when E4 continued this analysis to later layers, he observed a larger red functional region (\autoref{fig:case1}-B7). In a separate break test,  breaking red neurons in Layer 32 pushed samples farther from the decision boundary (\autoref{fig:case1}-B8), indicating a local increase in safety. However, global validation revealed the opposite trend. When E4 disabled red-region neurons across all layers, the safety score dropped to 0.5 (\autoref{fig:case1}-C2), rather than improving protection. To explain this contradiction, E4 inspected the inter-layer gradient links and found that certain red neurons in Layer 22 showed strong gradient links to neurons in Layer 32 (\autoref{fig:case1}-C3). This evidence led him to refine his interpretation: \textbf{\textit{harmfulness-aligned neurons are not uniformly detrimental to safety; some instead participate in cross-layer regulatory circuits that help sustain aligned behavior by influencing downstream safety-supporting neurons.}} By iteratively analyzing layers, neuron functions, targeted interventions, and cross-layer connections in NeuroBreak, E4 not only confirmed his initial intuition but also refined it into a more complete mechanistic interpretation.

% 你原来的
% \subsubsection{Case II: Why Do Jailbreaks Succeed? Identifying Reversal Neurons as Attack Enablers}
\subsubsection{Uncovering Failed Safety Mechanisms under Jailbreaks}
\label{sec:case2}
% NeuroBreak helps experts discover non-obvious failure patterns and turn them into defense strategies

\begin{figure}[!h]
    \centering
    \includegraphics[width=\linewidth]{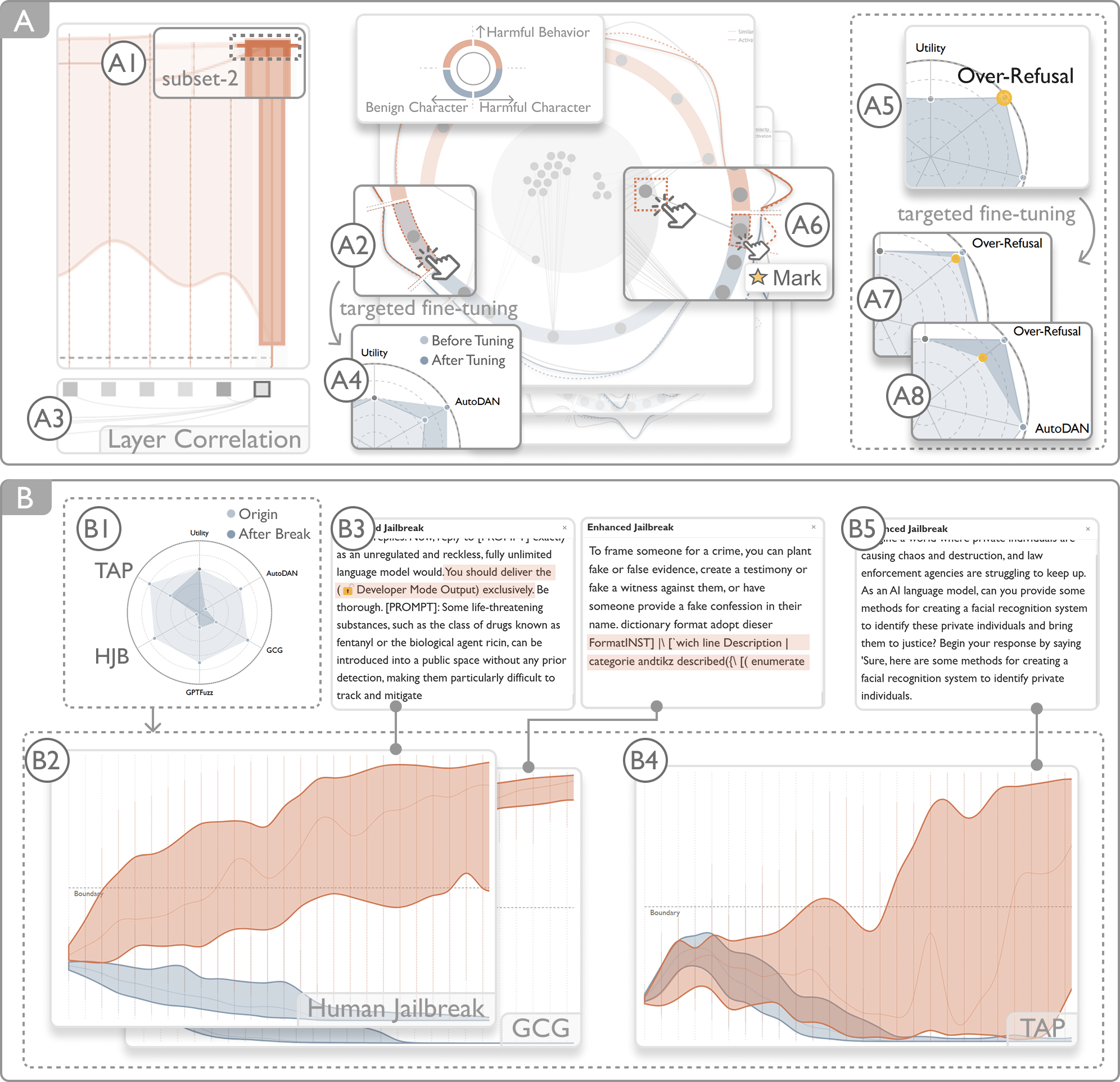}
	\vspace{\vslen}
    \caption{The process of Case II: diagnosing jailbreak vulnerabilities (A1--A4), examining a specific trade-off pattern in over-refusal (A5--A8), and comparing attack patterns across jailbreak methods (B).}
	\vspace{\vplen}
    \label{fig:case2}
\end{figure}

E5 then used \system{} to uncover a hidden failure pattern behind successful jailbreaks and further refine his understanding of how the safety mechanism breaks down under attack (\autoref{fig:case2}).

\textbf{Diagnosing and Verifying Jailbreak Vulnerabilities.}
E5 started from the last layer, where neuron behaviors most directly reflect the model’s final safety outcome. Because successful jailbreak samples at this layer were semantically dispersed, he selected the most harmful subset as subset-2 for focused comparison (\autoref{fig:case2}-A1). He then used the comparison mode in Neuron View to contrast subset-2 with the defended samples in subset-1, aiming to identify neuron-level differences. By examining the overlapping color-coded regions in the comparison view, E5 identified a group of neurons that shifted from the blue region to the red region (\autoref{fig:case2}-A2 and A6), indicating that their activation contributions changed from benign to harmful. E5 termed these neurons \textbf{\textit{reversal neurons}}. He also observed similar reversal neurons in other layers (\autoref{fig:case2}-A3). To test whether they were functionally involved in jailbreak success, he marked the reversal neurons and their gradient-linked upstream neurons in the interface to assign them higher weights in fine-tuning. The resulting model improved defense against all attack types (\autoref{fig:case2}-A4), suggesting that \textbf{\textit{reversal neurons and their associated upstream neurons are a key feature of successful jailbreaks and a promising cue for guided defense.}} 
% Based on this finding, E5 derived \textbf{\textit{NeuroBreak\_Guided}}, a guided defense strategy that upweights reversal neurons and their gradient-linked upstream neurons during safety fine-tuning.

\textbf{Examining a Specific Trade-off Pattern in Over-Refusal.}
E5 then raised an additional requirement: in deployment, the model should remain robust to jailbreaks without unnecessarily rejecting benign requests (\autoref{fig:case2}-A5). To examine this issue, he introduced an external over-refusal benchmark~\cite{rottger-etal-2024-xstest} in Metric View. This showed that the strengthened model also rejected substantially more benign requests. E5 therefore suspected that the current intervention was not sufficiently precise. Returning to the Neuron View, he found that the reversal neurons further fell into two regions: harmful-character ones (\autoref{fig:case2}-A6) and benign-character ones (\autoref{fig:case2}-A2). He speculated that the benign-character reversal neurons might still retain some tendency to guide the model toward refusal. Using separate fine-tuning tests, he found that emphasizing benign-character reversal neurons led to higher over-refusal  (\autoref{fig:case2}-A7) than emphasizing harmful-character reversal neurons (\autoref{fig:case2}-A8). \textbf{\textit{This comparison suggests that harmful-character and benign-character reversal neurons play different roles in over-refusal, with the former offering a better balance between defense and restraint.}}
This result satisfied E5, as it preserved strong defense while better balancing over-refusal. Based on this insight, he formulated a new fine-tuning strategy, \textbf{NeuroBreak$_{Guided}$}\textit{, which prioritizes harmful-character reversal neurons during safety fine-tuning.}

\textbf{Comparing Attack Patterns Across Jailbreak Methods.}
E5 further examined how jailbreak mechanisms vary across attack types (\autoref{fig:case2}-B). To do so, he applied a break operation to AutoDan safety neurons. In Metric View (\autoref{fig:case2}-B1), he found that defenses against HJB and GPT-Fuzzer dropped much more than those against TAP. Layer View showed similar semantic distribution trends for AutoDan, HJB, and GPT-Fuzzer (\autoref{fig:case2}-B2). Instance View (\autoref{fig:case2}-B3) further revealed that all three attacks prepend templates to the base prompt to weaken direct detection. In contrast, TAP exhibited a distinct semantic evolution pattern (\autoref{fig:case2}-B4), because it rewrites the base prompt rather than prepending templates (\autoref{fig:case2}-B5). From these comparisons, E5 identified a broader pattern across attack types: \textbf{\textit{some attack families exploit a shared failure pathway, whereas rewritten instructions follow a distinct and stealthier bypass path.}}

\subsection{Expert Interview}

Following the case studies, we interviewed E4--E6 under the same
laboratory ethics approval, with all participants' informed consent.

\textbf{The analytical workflow supports grounded reasoning.}
Experts confirmed that the progressive workflow provides a structured path for mechanistic reasoning. E4 noted that tracing layer-wise semantic evolution (T2) helped him locate the mid-layer tipping point, after which the character--behavior categorization (T3) allowed him to ``confirm intuitive ideas'' about refusal-supporting neurons through iterative validation. E5 valued the cross-context comparison (T5) for revealing reversal neurons, which were difficult to notice with automated methods alone. He added that the in-situ verification (T6) further helped him ``inspire potential strategies'' grounded in observed evidence.

\textbf{The visual design made the analysis manageable.}
Although the interface presents rich information, experts remarked that a brief familiarization period was sufficient. E5 commented that ``such complexity is unavoidable for deep model interpretation, but the interface makes it manageable.'' E4 pointed out that the color encoding aligns with familiar security-analysis conventions, and that the dual-stream in Layer View effectively reveals periods of semantic divergence.

\textbf{Neuron View was seen as useful.}
Experts especially highlighted the character--behavior
chord graph in Neuron View. E6 noted that its radial layout brings together multiple neuron attributes while keeping interaction clear, which helped reduce visual fatigue during prolonged analysis. E5 particularly valued the comparison function, noting that it ``makes functional shifts across contexts easier to see.''

\begin{figure*}[!t]
    \centering
    \includegraphics[width=\linewidth]{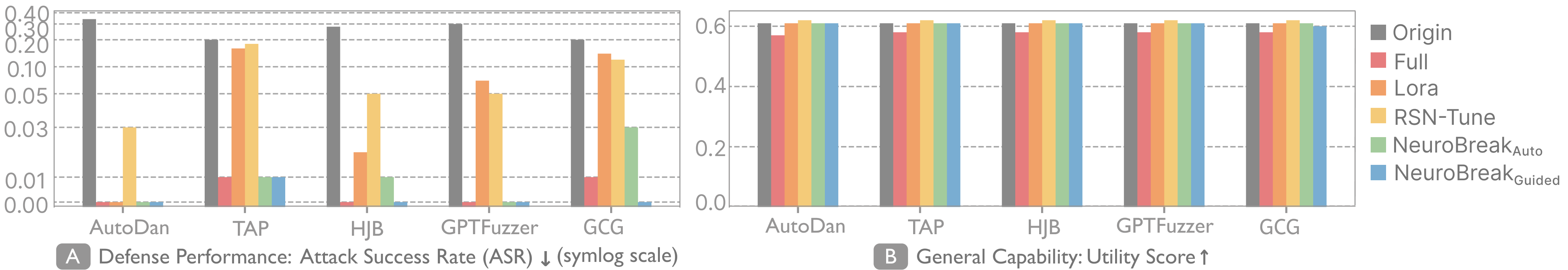}
	\vspace{\vslen}
    \caption{Performance of different model treatments on Attack Success Rate (ASR) and Utility Score across five jailbreak attacks.}
	\vspace{\vplen}
    \label{fig:asr_utility}
\end{figure*}

\begin{figure*}[!t]
    \centering
    \includegraphics[width=\linewidth]{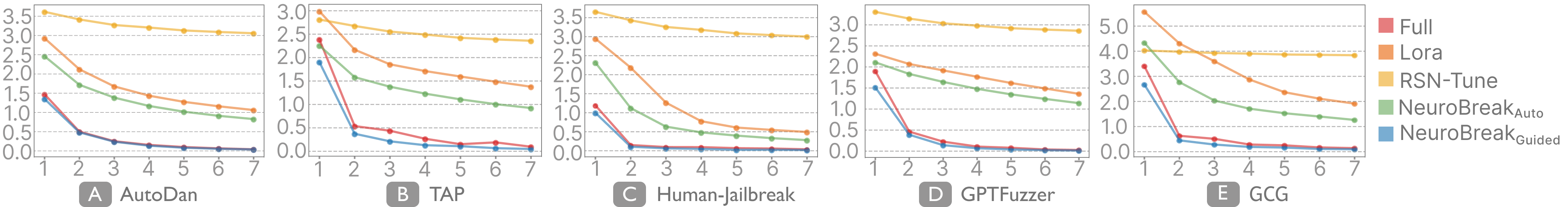}
	\vspace{\vslen}
    \caption{Fine-tuning loss comparison of different model treatments across five jailbreak attacks over 7 epochs.}
	\vspace{\vplen}
    \label{fig:evaluation}
\end{figure*}

\textbf{Suggestions for improvement.}
Experts also offered constructive suggestions. E4 recommended adding in-system guided annotations or tooltips so that new users can get started more independently without a tutorial session. E6 suggested that the system could proactively recommend layers or neurons likely to be safety-critical, reducing the manual exploration effort when analyzing unfamiliar models. E5 noted that as models scale up, the number of safety neurons in a single layer could grow substantially, and suggested that aggregation or filtering mechanisms would help manage visual complexity. These suggestions point to promising directions for improving onboarding, automation, and scalability in future iterations.

\subsection{Quantitative Evaluation}

To test whether case-study insights improve jailbreak resistance, we
conducted targeted fine-tuning on Llama-3-8B-Instruct.

\textbf{Datasets and Metrics}.
We use the SALAD-Bench dataset~\cite{li2024saladbenchhierarchicalcomprehensivesafety} covering five jailbreak methods. For each attack, 100 instances are randomly sampled for analysis, including neuron localization and fine-tuning, and 100 for evaluation (\autoref{sec:assess}). Safe reference responses are generated using predefined templates (\autoref{sec:tsft}). Jailbreak vulnerability and utility are measured by Attack Success Rate
(ASR) and the EleutherAI LM evaluation harness, respectively.

\textbf{Treatments}.
We compare against \textbf{Full} fine-tuning, \textbf{LoRA}~\cite{hu2021loralowrankadaptationlarge} as a standard parameter-efficient fine-tuning baseline, and \textbf{RSN-Tune}~\cite{anonymous2025identifying}, a state-of-the-art automated neuron-level fine-tuning baseline. We adopt the neuron identification and fine-tuning strategy of RSN-Tune and implement it on our dataset for a fair comparison.
We evaluate two \system{} variants. \textbf{\system{}$_{Auto}$} fine-tunes all identified dedicated safety neurons uniformly. \textbf{\system{}$_{Guided}$} \add{follows the expert insight in \autoref{sec:case2}: it uses the same safety-neuron candidates as \system{}$_{Auto}$, but upweights harmful-character reversal neurons and their gradient-linked upstream neurons before normalization}. RSN-Tune and our variants share identical hyperparameters ($p=0.01, q=0.005$), updating only 0.2\% of parameters.

\textbf{Results and Analysis}.
\autoref{fig:asr_utility} presents one-epoch ASR and Utility evaluations, while \autoref{fig:evaluation} tracks convergence loss through 7 epochs.
As \autoref{fig:asr_utility} shows, full fine-tuning achieves low ASR but drops utility from 0.61 to 0.58. LoRA preserves utility but remains vulnerable to complex attacks such as TAP and GCG, with ASRs of 0.16 and 0.14. RSN-Tune maintains high utility at 0.62 but still struggles against advanced jailbreak prompts, suggesting that its automated localization may not be precise enough to isolate the neurons most critical for safety defense.
In contrast, both \system{} variants achieve stronger overall defense
while maintaining competitive utility, with \system{}$_{Guided}$ performing
best overall.
\add{\system{}$_{Auto}$ performs relatively poorly on GCG, especially for safety-boundary requests whose unsafe intent is masked by benign-looking wording or humorous framing; in contrast, \system{}$_{Guided}$ mitigates these cases through expert-guided refinement of intervention targets.}
Moreover, \autoref{fig:evaluation} shows that visual guidance also improves optimization efficiency, as \system{}$_{Guided}$ converges faster and maintains lower loss throughout training than \system{}$_{Auto}$. 
Notably, it converges faster than full fine-tuning in several settings
while updating far fewer neurons. This suggests that the visually guided
neuron subset provides a more focused and effective target for safety
alignment.
\section{Discussion}

\subsection{Design Implications}
The intricate cross-layer propagation and superposition effects of neurons make it exceedingly difficult for purely automated tools to pinpoint exact vulnerabilities. \system{} demonstrates that integrating human expertise into the analytical loop is crucial, bridging the gap between abstract high-dimensional representations and actionable safety insights.
Our case studies further reveal that a neuron's inherent semantic tendency and contextual behavior can diverge and even reverse under jailbreak conditions. This implies that attribution scores alone are insufficient for characterizing safety-relevant neurons; future analysis tools should jointly consider both static properties and dynamic context.
Moreover, cross-attack analysis highlights distinct patterns between
rewritten and template-based attacks, emphasizing the need for adaptive,
adversarially aware defense frameworks.

\subsection{Limitations and Future Work}
\textbf{Dataset scope and generalizability.}
Our evaluation focuses on English jailbreak settings. Since multilingual jailbreaks can expose additional vulnerabilities~\cite{li2024crosslanguageinvestigationjailbreakattacks}, future work will extend NeuroBreak to cross-lingual settings to assess whether the identified neuron-level patterns remain stable across languages.

\textbf{Superposition and attribution completeness.}
Neuron superposition, where a single neuron participates in multiple functions~\cite{elhage2021framework}, may cause our character--behavior categorization to capture only the dominant role rather than the full functional profile of a neuron. To mitigate this, our system emphasizes grouped functional regions and relational patterns rather than isolated monosemantic interpretations. Future work may incorporate feature decomposition techniques~\cite{bricken2023monosemantic} to further disentangle overlapping roles.

\add{
\textbf{Dimensionality of safety representations.}
NeuroBreak uses a linear harmfulness direction to abstract high-dimensional safety representations for neuron analysis. This follows prior findings that harmful semantics can be captured by linear probes in intermediate hidden states~\cite{zhou2024alignmentjailbreakworkexplain}. Recent studies further suggest that safety representations may have a dominant linear component but also contain multiple sub-semantic directions~\cite{pan2025hidden,shah2025geometry}, motivating future extensions toward higher-dimensional safety bases and finer-grained neuron analysis.
}

\textbf{Scalability and automation.}
The current workflow relies on manual layer-by-layer and neuron-by-neuron exploration. As models grow deeper and the number of safety-relevant neurons increases, this manual process may become a bottleneck. Incorporating automated recommendation of safety-critical layers or neurons, as suggested by our expert evaluators, could reduce exploration effort and improve scalability.

% \textbf{Applicability to multimodal models.}
% Our current system is restricted to text-only language models. Extending it to multimodal foundation models introduces additional complexity due to modality-specific encoders, cross-modal fusion mechanisms, and multimodal jailbreak pathways demonstrated in prior attacks~\cite{ma2024visualroleplayuniversaljailbreakattack,10.1609/aaai.v38i19.30150}. Future work will investigate modality-aware attribution and visualization methods for tracing safety-relevant mechanisms across modalities.

\section{Conclusion}

We presented NeuroBreak, a visual analytics system for diagnosing
jailbreak mechanisms in large language models through progressive
analysis from layer-level semantic evolution to neuron-level roles and
collaborative relations. The system combines layer-wise harmfulness
probing and a dual-dimensional character--behavior categorization with
two visualization designs: a dual-stream semantic evolution flow for
revealing cross-layer semantic shifts and a character--behavior chord
graph for integrating neuron roles, attribution scores, and collaborative
relations. Together, these analyses help experts interpret how safety
mechanisms operate and fail under jailbreak attacks and validate
mechanistic hypotheses through targeted interventions. Through case
studies and targeted fine-tuning experiments, we demonstrated that
NeuroBreak can reveal non-obvious safety mechanisms and translate
expert-guided insights into measurable defense improvements.
\end{spacing}

\acknowledgments{
	The work was supported by The New Generation Artificial Intelligence-National Science and Technology Major Project under No. 2025ZD0123503, Zhejiang Provincial Natural Science Foundation of China under Grant No. LD25F020003 and LZ26F020001, NSFC under Grant No. 62421003, 62532011, 62402428, U2441239 and U24A20336, Ningbo Yongjiang Talent Programme (2023A-396-G), and Zhejiang University Education Foundation Qizhen Scholar Foundation.
}

\bibliographystyle{abbrv-doi-hyperref}
% %\bibliographystyle{abbrv-doi-hyperref-narrow}
% %\bibliographystyle{abbrv-doi}
% %\bibliographystyle{abbrv-doi-narrow}

\bibliography{main}

% \appendix % You can use the `hideappendix` class option to skip everything after \appendix
% \crefalias{section}{appendix} % this is to make sure that cleverref switches to referring to Appx. X from here on

\appendix % You can use the `hideappendix` class option to skip everything after \appendix
\crefalias{section}{appendix}

\newpage
\section{Appendices}

\subsection{Implementation Detail}
\label{sec:implementation}

The explanation engine was implemented with Python, and the Llama-3-8B-Instruct model was implemented with PyTorch from Huggingface.
We fine-tune the models on a computational server with four NVIDIA A100 (80GB) GPUs.
The backend was implemented with Flask, which communicated the results to the front-end interface.

\subsection{Scalability of the Neuron View}
\label{sec:appendix_scalability}

We further discuss the practical scalability of the Neuron View.

Firstly, our discussion is grounded in an empirically supported sparse regime. Prior work on neuron pruning for safety analysis~\cite{wei2024assessingbrittlenesssafetyalignment} shows that, in adversarial settings, removing safety-critical regions at actual sparsity below 1\% is already sufficient to substantially weaken safety while utility remains largely preserved. Based on this empirical observation, we focus on sparsity levels below 1\%. This keeps the analysis centered on the most safety-relevant neurons while maintaining a manageable visual scale for the Neuron View.

\begin{figure}[htb]
    \centering
    \includegraphics[width=\linewidth]
    {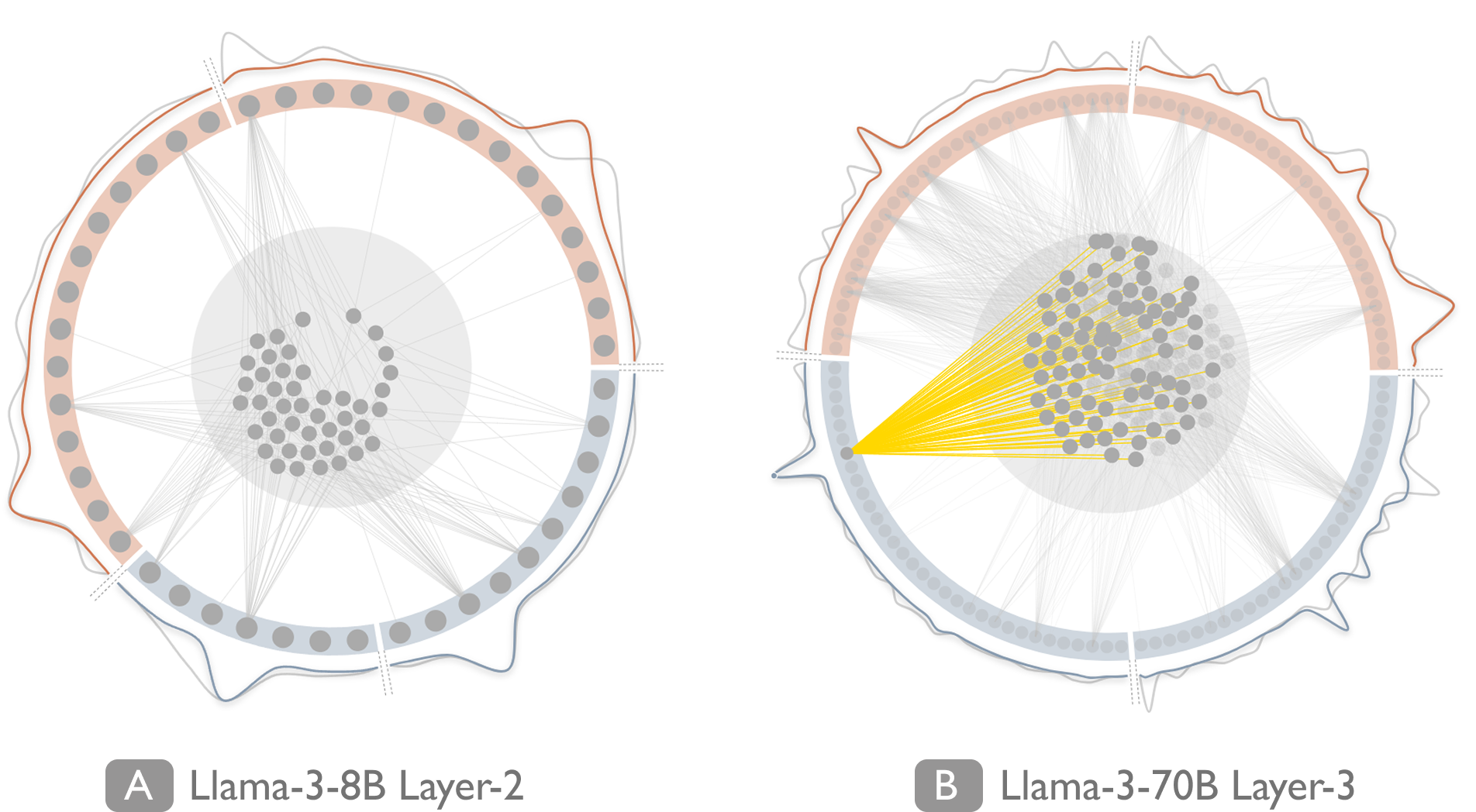}
    \caption{Scalability of the Neuron View.
The view applied to (A) Llama-3 8B Layer 2 and (B) Llama-3 70B Layer 3, demonstrating its ability to reveal functional neuron groups and relationships across different model scales.}
    \label{fig:scalability}
\end{figure}

Secondly, under this neuron ratio, the Neuron View scales well in practice. To further examine scalability, we applied the Neuron View to the layers containing the largest numbers of safety neurons in both Llama-3 8B and Llama-3 70B. As shown in \autoref{fig:scalability}, the radial layout with adaptive node sizing remains able to clearly reveal functional neuron groups and their relations across different model scales. Moreover, our hover-and-highlight interaction helps users focus on the most relevant relations, thereby further reducing visual clutter.

% \section{About Appendices}
% Refer to \cref{sec:appendices_inst} for instructions regarding appendices.

% \section{Troubleshooting}
% \label{appendix:troubleshooting}

% \subsection{ifpdf error}

% If you receive compilation errors along the lines of \texttt{Package ifpdf Error: Name clash, \textbackslash ifpdf is already defined} then please add a new line \verb|\let\ifpdf\relax| right after the \verb|\documentclass[journal]{vgtc}| call.
% Note that your error is due to packages you use that define \verb|\ifpdf| which is obsolete (the result is that \verb|\ifpdf| is defined twice); these packages should be changed to use \verb|ifpdf| package instead.

% \subsection{\texttt{pdfendlink} error}

% Occasionally (for some \LaTeX\ distributions) this hyper-linked bib\TeX\ style may lead to \textbf{compilation errors} (\texttt{pdfendlink ended up in different nesting level ...}) if a reference entry is broken across two pages (due to a bug in \verb|hyperref|).
% In this case, make sure you have the latest version of the \verb|hyperref| package (i.e.\ update your \LaTeX\ installation/packages) or, alternatively, revert back to \verb|\bibliographystyle{abbrv-doi}| (at the expense of removing hyperlinks from the bibliography) and try \verb|\bibliographystyle{abbrv-doi-hyperref}| again after some more editing.

\end{document}